\documentclass{jfm}
\usepackage{graphicx}
\usepackage{bm}
\usepackage{color}
\usepackage{subfigure}
\usepackage{natbib}
\usepackage{amssymb}
\usepackage{epstopdf}
\def\drawline#1#2{\raise 2.5pt\vbox{\hrule width #1pt height #2pt}}
\def\spacce#1{\hskip #1pt}
\def\solid{\drawline{24}{.5}\nobreak\ }
\def\bdash{\hbox{\drawline{4}{.5}\spacce{2}}}
\def\dashed{\bdash\bdash\bdash\bdash\nobreak\ }
\def\bdot{\hbox{\drawline{1}{.5}\spacce{2}}}
\def\dotted{\hbox{\leaders\bdot\hskip 24pt}\nobreak\ }
\def\chndot{\hbox {\drawline{9.5}{.5}\spacce{2}\drawline{1}{.5}\spacce{2}\drawline{9.5}{.5}}\nobreak\ }
\def\chndotdot{\hbox {\drawline{8}{.5}\spacce{2}\drawline{1}{.5}\spacce{2}\drawline{1}{.5}\spacce{2}\drawline{8}{.5}}\nobreak\ }

\begin{document}

\title[Sidewall effects in Rayleigh-B\'enard convection] {Sidewall effects in Rayleigh-B\'enard convection}

\author[Richard J.A.M. Stevens, Detlef Lohse and Roberto Verzicco]
{Richard J.A.M. Stevens$^{1,2}$,  Detlef Lohse$^1$ and Roberto Verzicco$^{1,3}$}

\affiliation{
$^1$Department of Science and Technology and J.M. Burgers Center for Fluid Dynamics, University of Twente, P.O Box 217, 7500 AE Enschede, The Netherlands,\\
$^2$Dept.\ of Mech.\ Engineering, Johns Hopkins University, Baltimore, Maryland 21218, USA\\
$^3$Dept. of Ind. Eng., Universit\`a di Roma "Tor Vergata",Via del Politecnico 1, 00133, Roma.}

\maketitle

\begin{abstract}
We investigate the influence of the temperature boundary conditions at the sidewall on the heat transport in Rayleigh-B\'enard (RB) convection using direct numerical simulations. For relatively low Rayleigh numbers $Ra$ the heat transport is higher when the sidewall is isothermal, kept at a temperature $T_c+\Delta/2$ (where $\Delta$ is the temperature difference between the horizontal plates and $T_c$ the temperature of the cold plate), than when the sidewall is adiabatic. The reason is that in the former case part of the heat current avoids the thermal resistance of the fluid layer by escaping through the sidewall that acts as a short-circuit. For higher $Ra$ the bulk becomes more isothermal and this reduces the heat current through the sidewall. Therefore the heat flux in a cell with an isothermal sidewall converges to the value obtained with an adiabatic sidewall for high enough $Ra$ ($\simeq 10^{10}$). However, when the sidewall temperature deviates from $T_c+\Delta/2$ the heat transport at the bottom and top plates is different from the value obtained using an adiabatic sidewall. In this case the difference does not decrease with increasing $Ra$ thus indicating that the ambient temperature of the experimental apparatus can influence the heat transfer. A similar behavior is observed when only a very small sidewall region close to the horizontal plates is kept isothermal, while the rest of the sidewall is adiabatic. The reason is that in the region closest to the horizontal plates the temperature difference between the fluid and the sidewall is highest. This suggests that one should be careful with the placement of thermal shields outside the fluid sample to minimize spurious heat currents. When the physical sidewall properties (thickness, thermal conductivity and heat capacity) are considered the problem becomes one of conjugate heat transfer and different behaviors are possible depending on the sidewall properties and the temperature boundary condition on the `dry' side. The problem becomes even more complicated when the sidewall is shielded with additional insulation or temperature-controlled surfaces; some particular examples are illustrated and discussed. It has been observed that the sidewall temperature dynamics not only affects the heat transfer but can also trigger a different mean flow state or change the temperature fluctuations in the flow and this could explain some of the observed differences between similar but not fully identical experiments.
\end{abstract}

\section{Introduction} \label{section_introduction}
The classical system to study turbulent heat transfer is Rayleigh-B\'enard (RB) convection, i.e. the motion of a fluid layer in a box heated from below and cooled from above (\cite{ahl09}). The system has many applications in atmospheric and environmental physics, astrophysics and process technology. The control parameters of the system are the Rayleigh number $Ra = \beta g \Delta L^3/(\kappa\nu)$, the Prandtl number $Pr = \nu /\kappa$ and the aspect ratio $\Gamma = D/L$. Here, $L$ and $D$ are the height and diameter of the fluid sample, $g$ the gravitational acceleration, $\Delta$ the temperature difference between the bottom and the top of the sample and $\beta$, $\nu$ and $\kappa$ the thermal expansion coefficient, the kinematic viscosity and the thermal diffusivity of the fluid, respectively. Nowadays most experimental and numerical results on the Nusselt number $Nu$, the dimensionless heat transfer, agree up to $Ra\approx 2\times 10^{11}$ and are in agreement with the description of the Grossmann--Lohse model (\cite{gro00,gro01,gro02,gro04,ste13}). However, for higher $Ra$ the situation is more complex.

\begin{figure}
\centering
\subfigure{\includegraphics[width=0.49\textwidth]{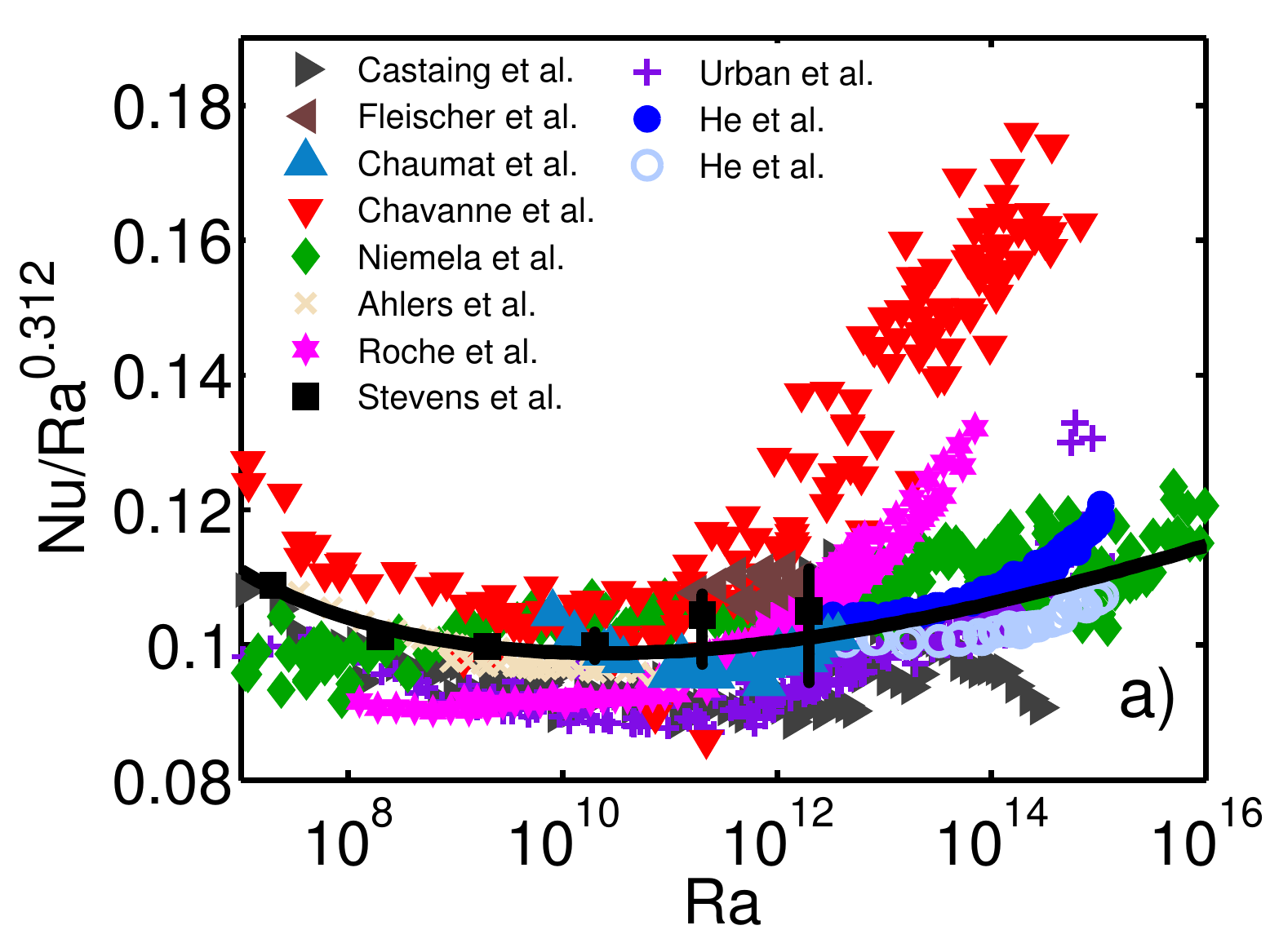}}
\subfigure{\includegraphics[width=0.49\textwidth]{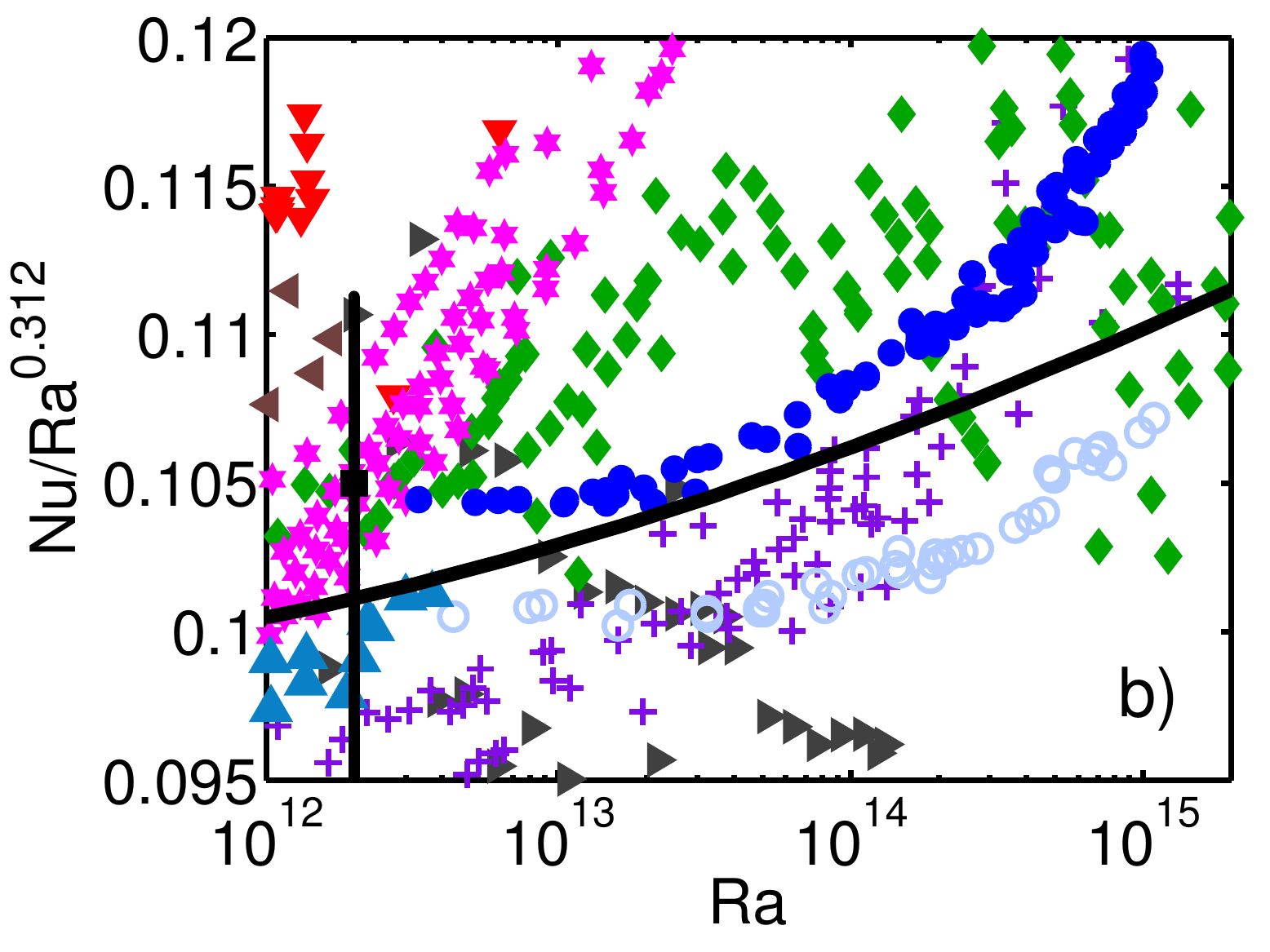}}
\caption{(Color online) (a) {$Nu$ versus $Ra$. Unless stated otherwise the data are for $\Gamma=1/2$.} The right--pointing triangles are the experimental data from \cite{cas89} with wall corrections. The experimental data are shown as stars \cite{roc10}, left--pointing triangles (\cite{fle02}), upward--pointing triangles (\cite{cha02}), downward--pointing triangles (\cite{cha01}), diamonds (\cite{nie00}), crosses (\cite{ahl09c}), hexagons (\cite{roc10}), plusses (\cite{urb11,urb12} {$\Gamma=1$}). The experimental results of \cite{he11} with $T_U-T_m \lesssim -3K$ are indicated by the filled (blue) circles and the results for $T_U-T_m \gtrsim +2K$ by the open (light blue) circles. The DNS results from \cite{ste10,ste10d} are indicated by the squares. (b) Zoom in on the high $Ra$ number regime.}
\label{fig:figure1}
\end{figure}

Most high-$Ra$ experiments are performed in samples with aspect ratio $\Gamma=1/2$ or smaller (for example $\Gamma=0.23$ in \cite{roc10}) owing to the dependence of $Ra$ on $L^3$ that, for a given volume of fluid, favors the increase of $L$ at the expense of $D$. Many of these experiments are performed with gaseous helium near its critical point (\cite{cas89,cha01,nie00,nie01,nie06,roc01,roc02,roc10,urb11,urb12}). More recently \cite{fun09}, \cite{ahl09c}, \cite{ahl09d}, \cite{ahl11} and \cite{he11} performed measurements at room temperature using highly pressurized gases. As is shown in figure \ref{fig:figure1} there are significant deviations among all experiments for $Ra\gtrsim 2\times 10^{11}$ 
and unfortunately there is no clear explanation for this disagreement. 

The studies of \cite{joh09} and \cite{ste10d} showed that the differences among the experiments cannot be explained by the fact that some setups use a constant heat flux condition at the bottom plate instead of a constant temperature condition, nor by the variations of $Pr$ (\cite{ste10d}). Recently, evidence has been found that suggests that part of the deviation might be related to the formation of different turbulent states in the high-$Ra$ number regime. Multiple states in RB convection were namely observed by \cite{roc02}, who found a bimodality of Nu with $7\%$ difference between the two data sets. Subsequently, \cite{chi04a} and \cite{sun05a} showed that a finite tilt of the sample can cause a transition between different flow states. Later \cite{xi08} and \cite{wei10b} found that in a $\Gamma=1/2$ sample the flow can be either in a single-roll state or in a double-roll state, each with a specific heat transport. Recently, \cite{nie10b} found two $Nu/Ra^{1/3}$ branches in a $\Gamma=1$ sample. The high Ra branch is $20\%$ higher than the low-$Ra$ branch. Very recently, Ahlers and coworkers (see \citealt{fun09}, \citealt{ahl09c,ahl09d,ahl11} and \citealt{he11}) found two different branches in one experiment. In these experiments the temperature difference between the average temperature inside the cell and the temperature outside the cell determines the state of the system. Also in the experiments of \cite{roc10} two different turbulent states were observed in a $\Gamma=0.23$ sample in the range $10^{12} \leq Ra \leq 10^{13}$. In addition, they showed that in the high $Ra$ number regime the heat transport can very strongly depend on the characteristics of the sidewall and the aspect ratio. Finally, \cite{poe11} showed, by two--dimensional RB simulations, that a different flow organization can lead to significant differences in the heat transport.

Due to technical difficulties, the physical properties and boundary conditions of the sidewall can only be controlled up to a certain degree in experiments. In addition, testing different sidewall configurations is a very time-consuming task as the entire sample has to be disassembled to replace the sidewall. On the other hand, direct numerical simulations (DNSs), even though they cannot reach as high $Ra$ numbers as obtained in some experiments, offer a good possibility to study the influence of the physical properties and the boundary conditions at the sidewall as they can be exactly controlled. By comparing the differences between simulation of a RB sample with an adiabatic sidewall and of RB samples with several other sidewall configurations we aim at getting a better understanding of the importance of sidewall effects. 

\begin{figure}
\centering
\subfigure{\includegraphics[width=0.54\textwidth]{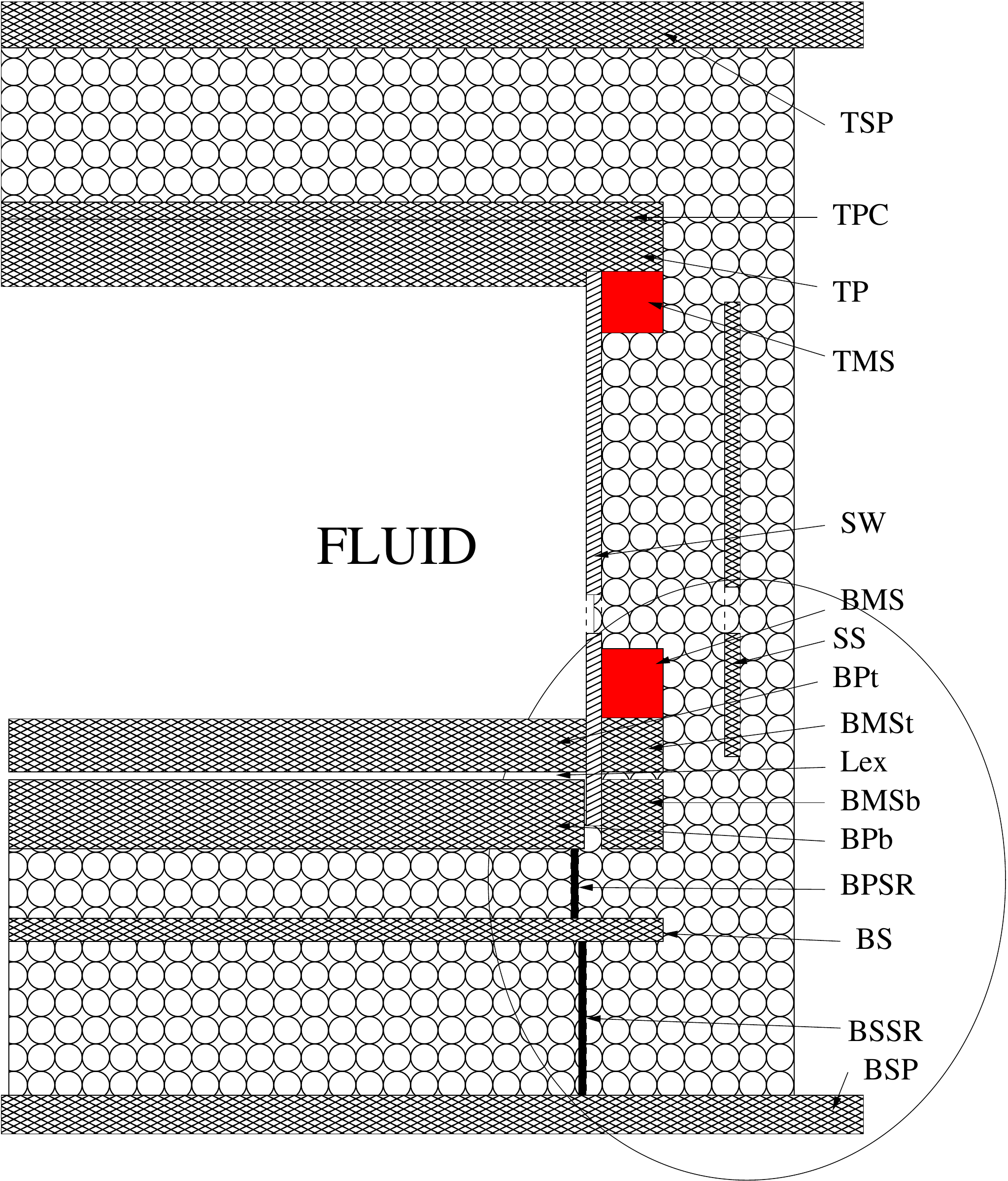}}
\subfigure{\includegraphics[width=0.44\textwidth]{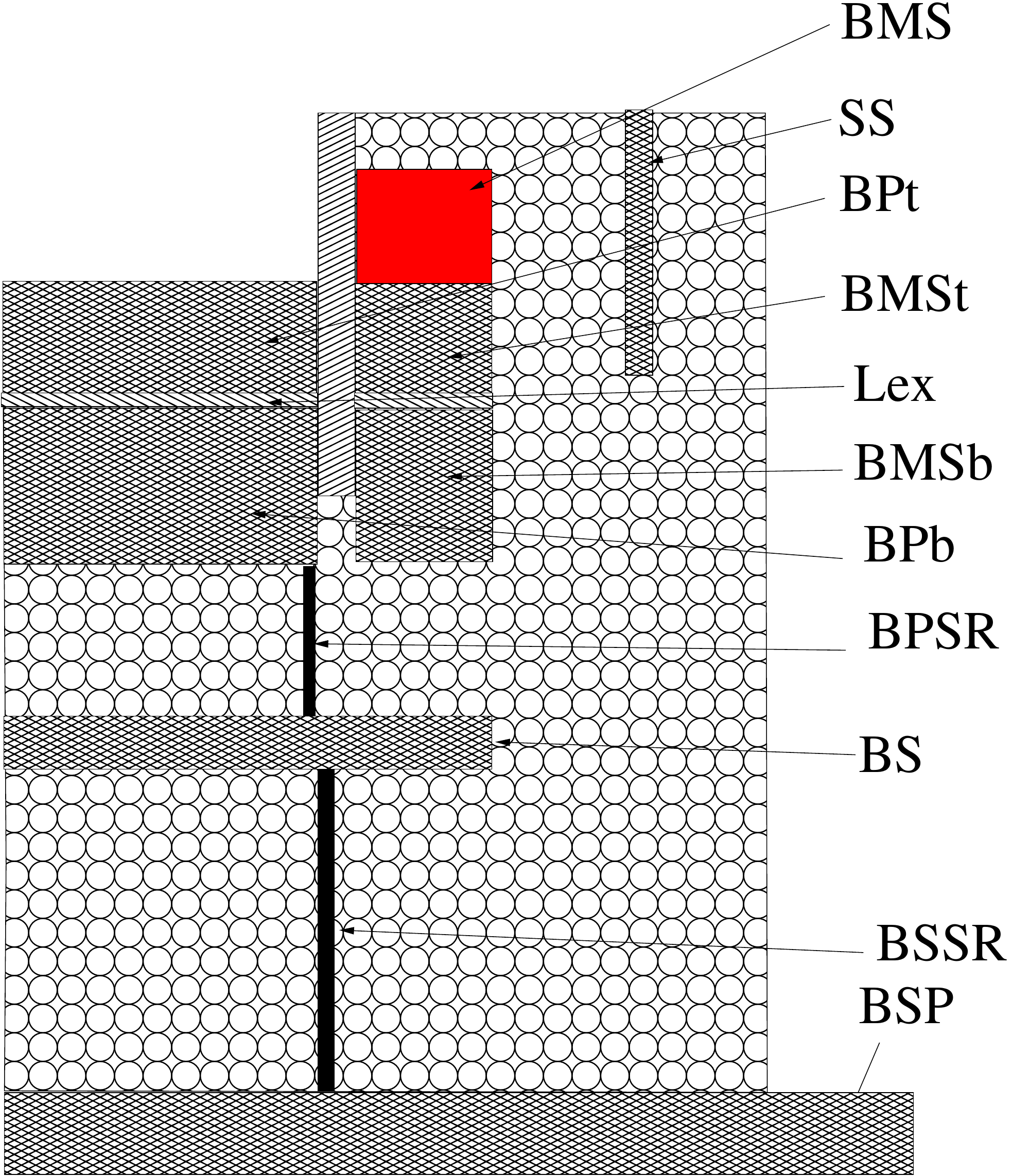}}
\caption{a) Schematic diagram of the G\"ottingen RB setup (adapted from \cite[]{ahl09d}). From bottom to top, we have the bottom support plate (BSP), the bottom-shield support ring (BSSR), the bottom shield (BS), the bottom-plate support ring (BPSR), the bottom plate bottom (BPb), the Lexan plate (Lex), the bottom plate top (BPt), the bottom micro-shield bottom (BMSb), the bottom micro-shield top (BMSt), the bottom micro-shield (BMS indicated in red), the side-shield (SS) with its water cooling coil (WCC), the Plexiglas sidewall (SW), the top micro-shield (TMS, indicated in red), the top plate (TP), the top-plate cover (TPC) and the top support plate (TSP). b) Enlargement of the bottom--plate/sidewall assembly.}
\label{fig:figure2}
\end{figure}

The influence of the sidewall on the heat transport has been investigated before by \cite{ahl00}, \cite{roc01b}, \cite{ver02}, and \cite{nie03}. The main conclusions are summarized in the review by \cite{ahl09}. In short, the phenomenological models of \cite{ahl00} and \cite{roc01b} showed that one cannot fully account for the effect of the sidewall by simply subtracting the corresponding heat transferred to an empty cell. This was confirmed by \cite{nie03} who simulated an idealized two--dimensional convection problem with a conducting side surface and a fly-wheel-like structure in the bulk in order to mimic the mean flow sweeping the walls. \cite{ver02} modeled the physical properties of the sidewall in three--dimensional DNSs and found that, for usual sidewall thicknesses, the heat traveling from the hot to the cold plates directly through the sidewall is negligible due to the heat exchanged at the fluid/wall interface. In contrast, the modified temperature boundary conditions alter the mean flow yielding significant $Nu$ number corrections in the low Ra number range. All these works suggested that the sidewall effects vanished for increasing $Ra$. However, recent experiments of \cite{roc10} and \cite{he11} indicate that the properties and temperature boundary conditions of the sidewall are also important at higher $Ra$. In this paper we will use DNSs to study the influence of the physical properties of the sidewall and the temperature boundary conditions at the sidewall on the heat transport and the flow dynamics in RB convection.

First, we will discuss the employed numerical procedure in section \ref{section_Numerical}. In section \ref{section_isothermal} we will show the difference between simulations with an adiabatic and an isothermal sidewall. This is an interesting comparison because in most experiments (\cite{bro05,sun05e,ahl09d,kun11}) the adiabatic temperature boundary condition is obtained by placing a sidewall temperature shield around the sidewall that is covered by a layer of insulation, see for example the sketch in figure \ref{fig:figure2}a of the RB sample used in the G\"ottingen experiments. The temperature of the side-shield is maintained at $T_M = T_c+\Delta/2$, where $T_c$ is the temperature of the top plate, since this coincides with the mean temperature of the fluid in the bulk $T_m$ (when the Oberbeck--Boussinesq approximation is fully valid). In figure \ref{fig:figure1}b it is shown that the heat transport that is measured in the G\"ottingen experiments depends on $T_U-T_m$, where $T_U$ is the temperature outside the RB sample. In order to investigate the possible influence of the temperature outside the cell we will consider different sidewall temperatures in section \ref{section_isothermal}. As a completely isothermal sidewall is an oversimplification of the experimental case we will consider the case in which only the bottom and top $1.5\%$ of the sidewall is kept at $T_M$ and the rest of the sidewall is adiabatic in section \ref{section_Mixed}. This case is based on the design of the G\"ottingen RB setup in which micro-shields with a temperature $T_M$ are placed just above (below) the lower (upper) plate in order to prevent that the insulated region between the sidewall and the side-shield is influenced by the temperature of the horizontal plates (see figure \ref{fig:figure2}). In order to come closer to the experimental situation we simulate the effect of the physical properties of the sidewall in section \ref{section_Physical} for some particular setups. In section \ref{Isolation layer} we will also consider the effect of thermal shields at a fixed temperature and an external layer of insulating foam where porous convection occurs. Finally, in section \ref{section_bimod} a brief account of the changes induced in the flow dynamics by the presence of a non--ideal sidewall is given. We will conclude the paper with a short summary and some closing remarks. 
An Appendix has been added at the end of the paper with two tables containing the most relevant results of all the numerical
simulations presented and discussed in this study.

\begin{figure}
\centering
\centering
\subfigure{\includegraphics[width=0.55\textwidth]{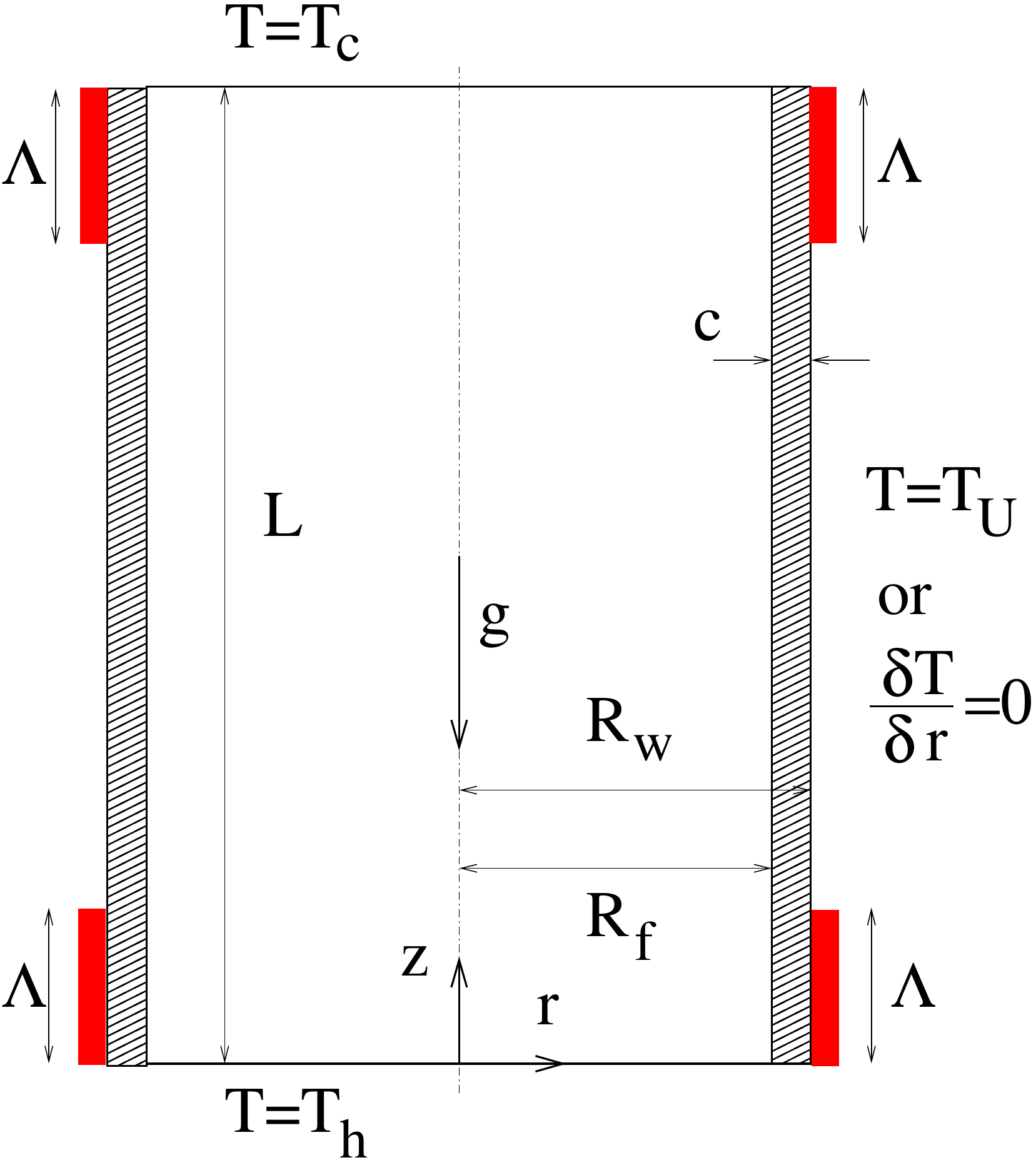}}
\caption{Sketch of the numerical setup of the problem: The upper and lower boundaries are isothermal and no--slip. The sidewall has a thickness $c = R_w-R_f$ and its 'dry' side at $r=R_w$ can be either adiabatic $\partial T/\partial r=0$ or isothermal at a temperature $T=T_U$; there is also the possibility to have isothermal boundary condition for the portions of the sidewall closest to the plates ($0 \leq z \leq \Lambda$ and $L-\Lambda \leq z \leq L$) while the rest of the sidewall is adiabatic. This will be referred to as 'mixed boundary condition'. The 'wet side' of the sidewall ($r=R_f$) is no--slip while its temperature is not a boundary condition but part of the solution of equation (\ref{eqt}) (conjugate heat transfer problem). When $c\neq 0$ the sidewall has thermal properties (density, specific heat and thermal conductivity) different from the fluid while for $c=0$ the sidewall is a simple boundary condition for velocity and temperature. Note that the hot and cold isothermal surfaces of the plates extend also below and above the 
sidewall when its thickness is not zero.  }
\label{fig:figure3}
\end{figure}

\section{Numerical procedure} \label{section_Numerical}
\noindent In order to simulate the physical properties of the sidewall we solve the non-dimensional Navier-Stokes equations within the Boussinesq approximation
\begin{eqnarray}
 \frac{D\textbf{u}}{Dt} = - \nabla P + \left( \frac{Pr}{Ra} \right)^{1/2} \nabla^2 \textbf{u} + \theta \textbf{$\widehat{z}$}, ~~~ \nabla \cdot \textbf{u} = 0 \mathrm{~~~~~~~~~~~~~ on ~~~~ V_f,} \label{eqt1}\\
\frac{D\theta}{Dt} = \frac{1}{(PrRa)^{1/2}} \frac{\rho_f C_{pf}}{\rho C} \nabla \cdot \left( \frac{\lambda}{\lambda_f} \nabla \theta \right) \mathrm{~~~~~~~~~~~~~~~~~~~~~~~~~~~ on ~~~~ V,}
\label{eqt}
\end{eqnarray}
\noindent where $V_f$ is the fluid domain $0\leq z \leq L$, $0\leq r \leq R_f$, and $0 \leq \phi \leq 2 \pi$ with $\phi$ the azimuthal coordinate, and $V$ ($0\leq z \leq L$, $0\leq r \leq R_w$, and $0 \leq \phi \leq 2 \pi$) the total domain (see figure \ref{fig:figure3}). 
{ $\rho$, $C$ and $\lambda$ are, respectively, density, specific heat and thermal conductivity and they assume the 
values of the fluid ($\rho_f$, $C_{pf}$ and $\lambda_f$ with $k_f=\lambda_f/(\rho_f C_{pf})$) or of the sidewall
($\rho_w$, $C_w$ and $\lambda_w$) depending on the specific point in the domain. }
Note that the physical properties of the sidewall are only incorporated in the simulations presented in sections \ref{section_Physical} to \ref{section_bimod}. In the simulations presented in sections \ref{section_isothermal} and \ref{section_Mixed} the physical properties of the sidewall are not taken into account by setting $R_f=R_w$, which sets the sidewall thickness to zero. It is worth mentioning that as the temperature field is solved on the whole domain $V$ no temperature boundary condition is required at the solid/fluid interface $r=R_f$. Instead the isothermal or adiabatic temperature boundary condition is imposed at the `dry' sidewall surface $r=R_w$. A third possibility is to have a `mixed temperature boundary condition' at the sidewall, i.e.\ isothermal for $0 \leq z \leq \Lambda$ and $L-\Lambda \leq z \leq L$ and adiabatic in between. This configuration mimics one particular feature of the apparatus of \cite{ahl09} as will be discussed later. 

As indicated in figure \ref{fig:figure3}, at the lower and upper plates, respectively, the temperatures $T_h$ and $T_c$
are prescribed so that they are modeled as isothermal surfaces. This implies that in our simulations the temperature
difference $\Delta$ is imposed and the heat flux entering the fluid $Q_f$  
is measured by the non dimensional Nusselt number 
$Nu = Q_fL/(\lambda_f \Delta)$  
\footnote{
In the present paper the heat flux $Q_f$ in the Nusselt number definition has been computed from the temperature gradient 
at the lower (hot) and upper (cold) plates $Q_f = \int_0^{2\pi}\int_0^{R_f} \lambda_f \nabla \theta \cdot {\bf n} {\rm d}S$. 
}.
This situation is similar to some of the high $Ra$ number experiments in which the temperature of the plates is kept constant (e.g. \cite{ahl09c,ahl09d,fun09}), while other high $Ra$ number experiments (\cite{nie00,nie01,nie06}) use an imposed heat flux at the lower plate and measure the temperature difference $\Delta$; also in this case the results are normalized through the Nusselt number. 
 Although we note that \cite{joh09} and \cite{ste10} have shown that, regardless the
imposition of $\Delta$ or $Q_f$, the same $Nu$ and the same $Nu$ versus $Ra$ relation is obtained (for high enough Rayleigh numbers) 
$Nu$ and $Q_f$ are not exactely the same quantity and changes in one do not necessarily imply changes also in 
the other.

It is worth mentioning that for the simulations with finite thickness sidewall the isothermal plates extend also below
and above the former (see figure \ref{fig:figure3}) therefore some heat is forced through the lateral wall even if only the 
flux entering and leaving the fluid layer is accounted for in the evaluation of the Nusselt number. 
This is trivial to achieve in numerical simulations by only considering the `wet' surfaces in the computation of $Nu$.
In contrast, in laboratory experiments the total heat entering the setup and the plates 
temperature difference are measured and disentangling the heat crossing only the fluid from the parasite currents is impossible.
An initial naive approach consisted of subtracting the heat current measured in an empty cell from that actually crossing the
setup with the fluid; this correction, however proved to be insufficient since it disregarded the conjugate heat transfer
between the fluid and the lateral wall (\cite{ahl00,roc01b}). Indeed numerical simulations by \cite{ver02} showed that some heat
was dynamically exchanged between fluid and sidewall when the latter had its own thermal properties and thickness and {\it ad hoc}
corrections were derived to properly account for this effect (\cite{ahl00,roc01b,ver02,nie03}).

In this paper the discussion of the results will always use the Nusselt numbers computed only for the fluid, although
some comments on the total heat entering the setup will be given in the Appendix.

In the equations (\ref{eqt1}-\ref{eqt}), \textbf{$\widehat{z}$} is the unit vector pointing in the opposite direction to gravity, $D/Dt = \partial_t + \textbf{u} \cdot \nabla $ the material derivative, $\textbf{u}$ the velocity vector with no-slip boundary conditions at all walls, and $\theta$ the non-dimensional temperature, $0\leq \theta \leq 1$. The equations have been made non-dimensional by using the length $L$, the temperature difference $\Delta$, and the free--fall velocity $U=\sqrt{\beta g \Delta L}$. The above equations have been written in a cylindrical coordinate frame and discretized on a staggered mesh by central second-order-accurate finite-difference approximations. The numerical method is described in detail by \cite{ver96}, \cite{ver97}, and \cite{ver02}. In this paper we present results for $2\times 10^6<Ra<2\times 10^{10}$ and $Pr=0.7$ (i.e., gas) in an aspect ratio $\Gamma=1/2$ sample.

In \cite{ste10} we investigated the resolution criteria that should be satisfied in a fully resolved DNSs and \cite{shi10} determined the minimal number of nodes that should be placed inside the boundary layers (BLs). In \cite{ste10d} we showed that a $769\times193\times769$ grid is sufficient to properly resolve a simulation at $Ra=2\times10^{10}$. Here we have used a resolution of $769\times257\times769$ for the simulations at $Ra=2\times10^{10}$. The increased number of radial nodes was used to have a proper resolution in the thermal BL that is formed along a isothermal sidewall or a boundary with physical properties. A proportionally increased radial resolution was used at lower Ra where simulations could be run to test the effects of the resolution on the heat transfer. For $2\times 10^7 \leq Ra \leq 2\times 10^9$ the flow could be simulated with over-resolved meshes (up to $50\%$ in each direction). These resolution tests always gave $Nu$ numbers within a few percent of the values obtained with the reference resolution and the difference decreased for increasing $Ra$. This observation gives us confidence that the results are reliable and can be used for the flow analysis. Finally, we emphasize that for lower $Ra$ (up to $Ra\approx 2\times10^8$) the results of our code agree well with completely independently written codes by \cite{shi09}, \cite{heb10}, and \cite{sch12}, and also agree well with experimental results.

Before starting the discussion of the results we wish to point out that although from figure \ref{fig:figure1} it is evident that the largest differences among the experiments show up for $Ra \geq 10^{11}$ running many simulations at these large Ra numbers is not feasible due to limitations of the computational resources. As a compromise we have restricted our investigation to the range $2\times 10^7 \leq Ra \leq 2\times 10^{10}$ and whenever possible we have exaggerated the non--standard features (temperature boundary conditions, wall thicknesses, etc.) in order to make their effects on the flow already visible at lower Ra.

\section{Isothermal sidewall} \label{section_isothermal}
\begin{figure}
\centering
\subfigure{\includegraphics[width=0.85\textwidth]{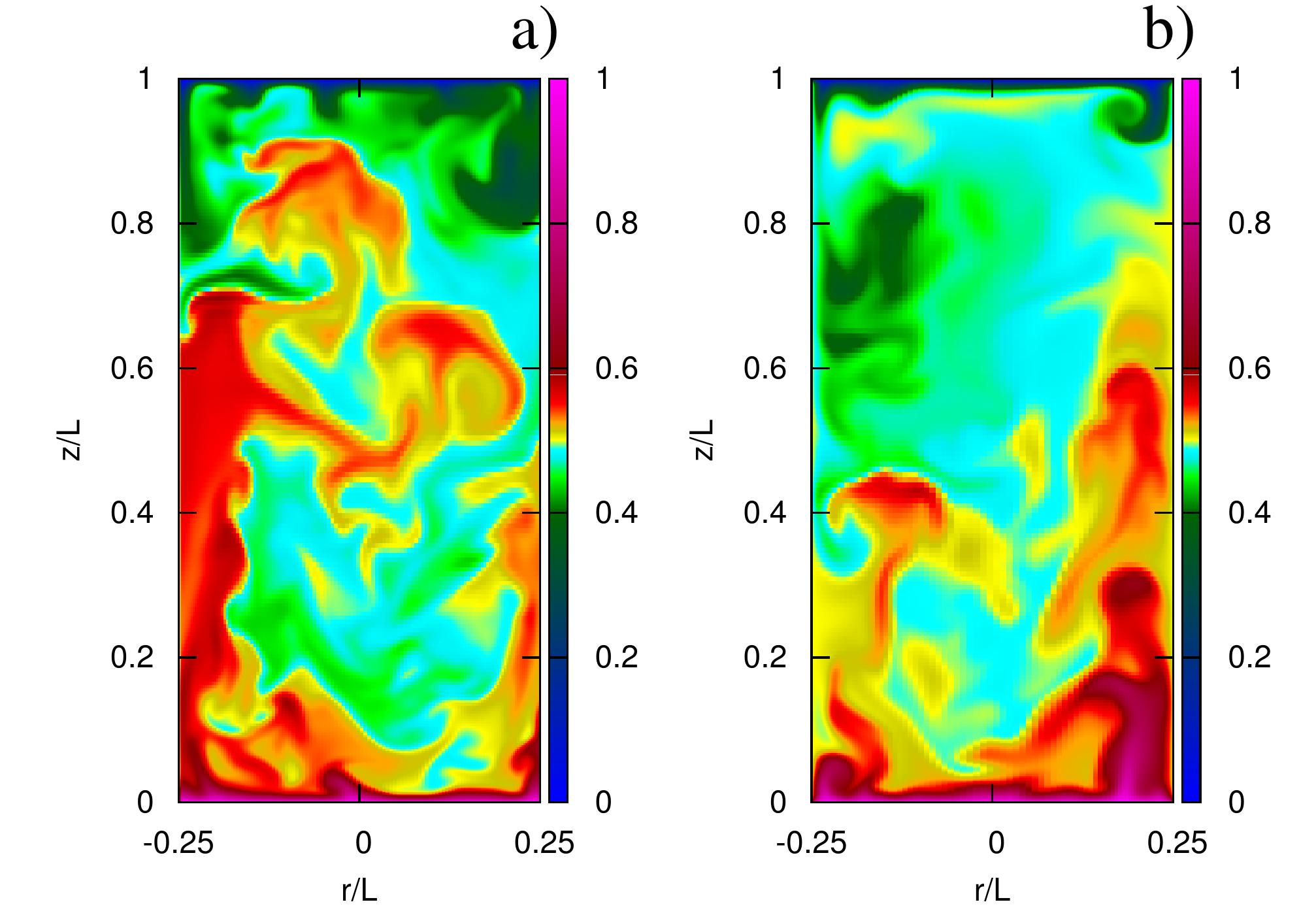}}
\caption{Visualization of instantaneous temperature field at $Ra=2\times10^8$ and $Pr=0.7$ in a $\Gamma=1/2$ sample with an (a) adiabatic and (b) an isothermal sidewall at $T_M$. Note the formation of the thermal BLs along the sidewall when the sidewall is isothermal.  }
\label{fig:figure4}
\end{figure}

\begin{figure}
\centering
\subfigure{\includegraphics[width=0.95\textwidth]{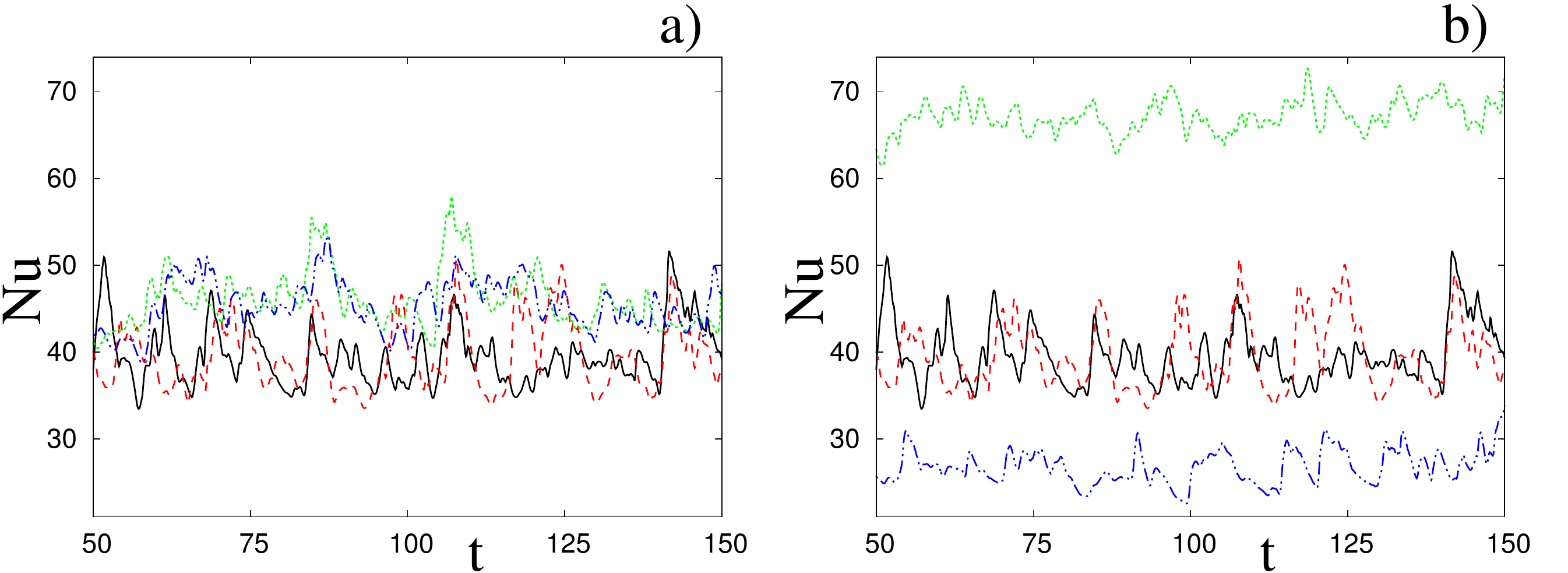}}
\caption{(Color online) $Nu$ number at the bottom (\chndotdot and blue) and top plate (\dotted and green) as function of time for $Ra=2\times10^8$ and $Pr=0.7$ in a $\Gamma=1/2$ sample when the temperature of the sidewall is kept at a) $T_U= T_c+\Delta/2=T_M$ and b) $T_U= T_c +0.75 \Delta$ ($T_U-T_m=0.071\Delta$). The heat transfer at the bottom (\solid and black) and top (\dashed and red) plates for a reference simulation with adiabatic sidewall is given in both panels.
The time $t$ is in non--dimensional units $L/U$.}
\label{fig:figure5}
\end{figure}

\begin{figure}
\centering
\subfigure{\includegraphics[width=0.95\textwidth]{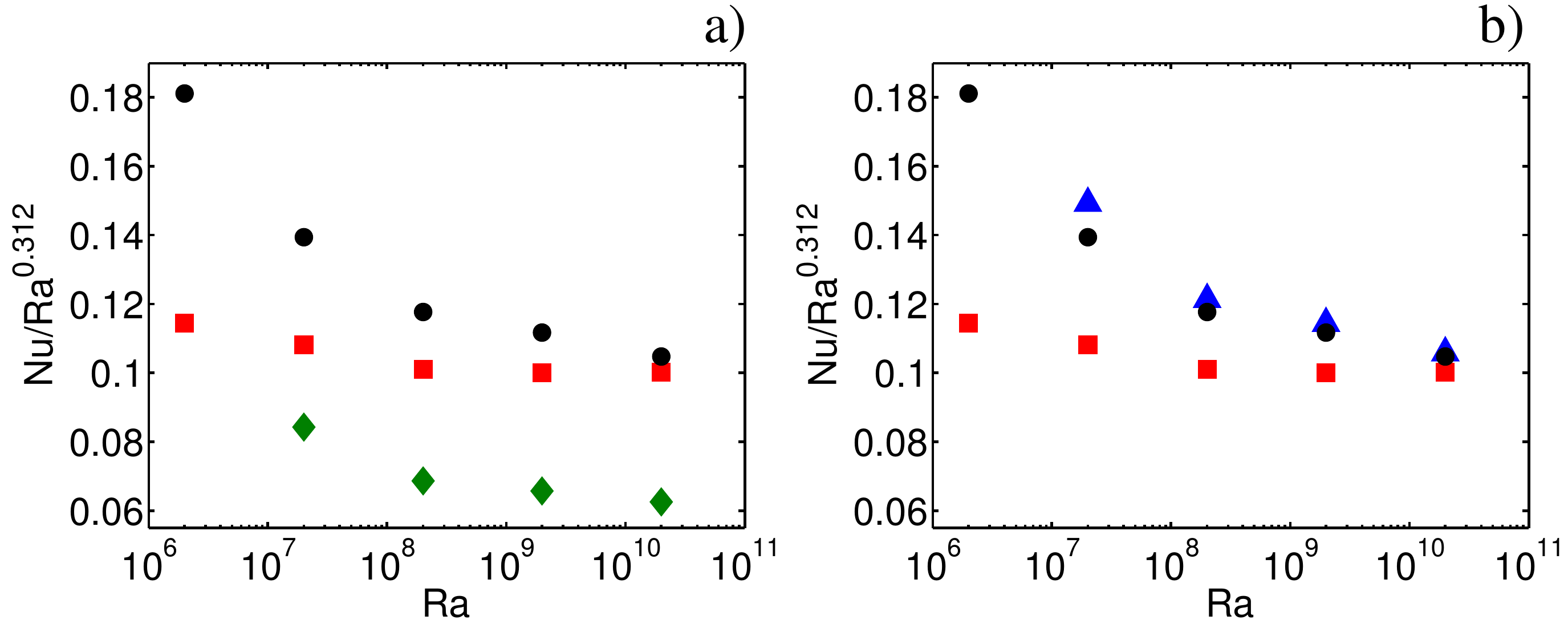}}
\caption{(Color online) The heat transport in a RB sample with an adiabatic sidewalls and a sidewall kept at $T_U=T_M$ is indicated by the (red) squares and (black) circles, respectively. Panel a) compares these results with simulations in which the sidewall temperature is $T_U = T_c+0.75 \Delta$ ($T_U-T_m=0.071\Delta$), which are indicated by (dark green) diamonds, and panel b) shows the results for the simulations in which the lower and upper $1.5 \%$ of the sidewall are kept at $T_U=T_M$ and the rest of the sidewall is adiabatic, which are indicated by (blue) triangles.
{ Note that up to $Ra=2\times 10^{10}$ the error bar 
of $Nu$ is smaller than the symbol size.}
}
\label{fig:figure6}
\end{figure}

\begin{figure}
\centering
\subfigure{\includegraphics[width=0.95\textwidth]{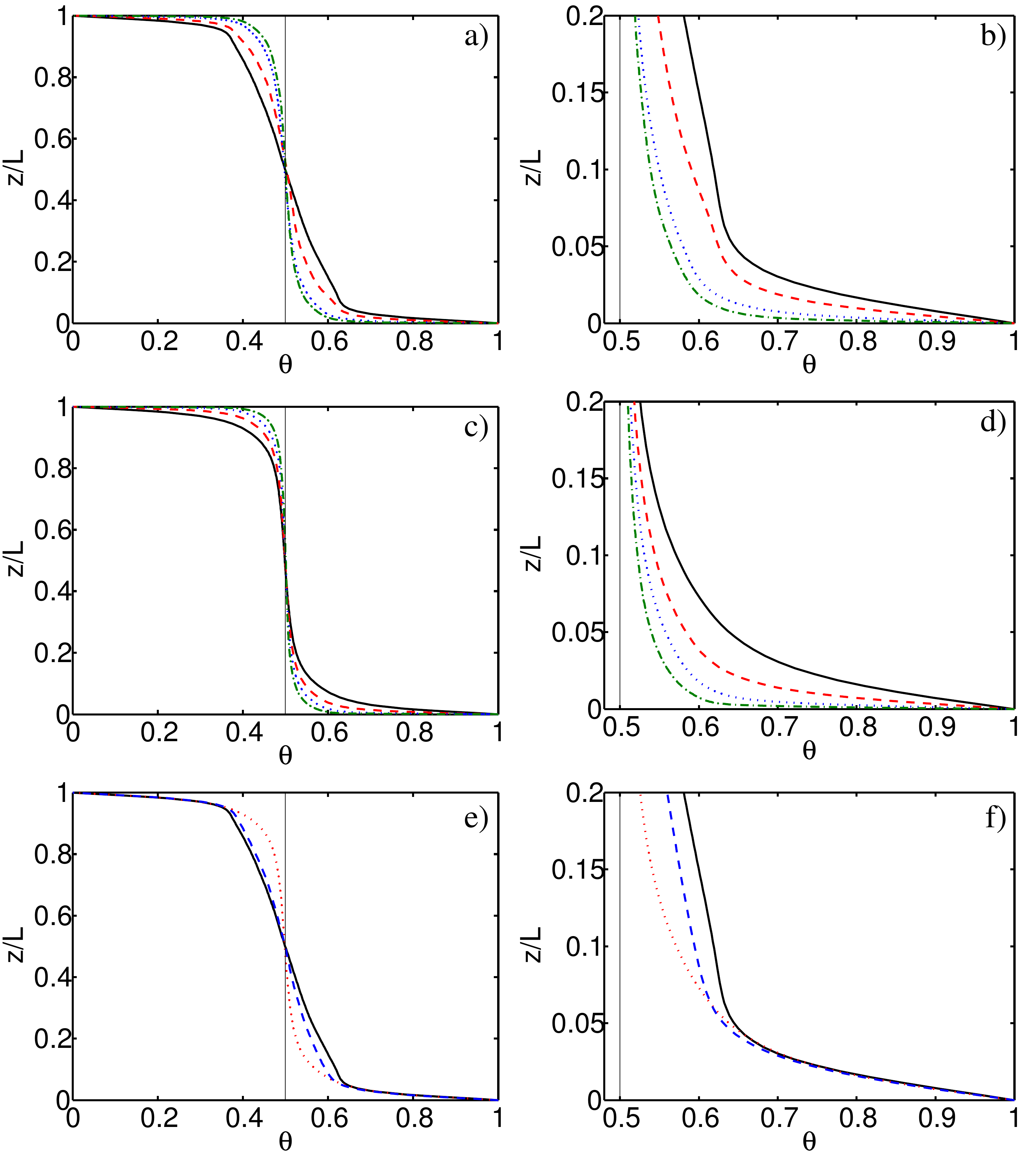}}
\caption{(Color online) The temperature profiles at $r=R-\delta_\theta$ as function of the height when the sidewall is adiabatic (top row, a) and b)) and when the sidewall is isothermal at $T_M$ (middle row, c) and d)) for different $Ra$ numbers: \solid (black) $Ra=2\times 10^7$, \dashed (red) $Ra=2\times 10^8$, \dotted (blue) $Ra=2\times 10^9$ and \chndot (green) $Ra = 2\times 10^{10}$. The bottom row (e) and f) compares the temperature profiles for an adiabatic sidewall (\solid black), a sidewall at $T_M$ (\dotted red), and the model sidewall ($T_M$ for $0 \leq z/L \leq 0.015$ and $0.985 \leq z/L \leq 1$ and adiabatic for $0.015 < z/L < 0.985$) (\dashed blue) at $Ra=2\times 10^7$. The plots on the left show the profiles over the entire domain while the plots on the right are for the region close to the bottom plate.}
\label{fig:figure7}
\end{figure}

Figure \ref{fig:figure4} shows a visualization of the instantaneous temperature field at $Ra=2\times 10^8$ in a $\Gamma=1/2$ sample when the sidewall is adiabatic and when it is kept at the constant temperature $T_M$. The figure shows that the difference between the two cases is the formation of a thermal BL along the sidewall when it is isothermal and this is most pronounced close to the horizontal plates. Figures \ref{fig:figure5}a and \ref{fig:figure6} show that at relatively low $Ra$ the heat flux is larger when the sidewall is isothermal than when the sidewall is adiabatic, even though the time-averaged heat flux through the entire sidewall is zero. However, locally there is a heat flux from the fluid to the sidewall in the lower half of the sample and vice versa in the top half. Because part of the heat current avoids the thermal resistance of the fluid in this way the heat transport measured at the horizontal plates is higher in the case of isothermal sidewalls. Figure \ref{fig:figure6} shows that the difference between the heat transport measured with an adiabatic sidewall and with a sidewall kept at $T_M$ decreases with increasing $Ra$. The reason is that with increasing $Ra$ the temperature becomes more isothermal in the bulk. 
Figure \ref{fig:figure7} confirms that the azimuthally and time averaged temperature close to the sidewall, more precisely at $r=R-\delta_\theta$, where $\delta_\theta = L/(2Nu)$ is the thermal BL thickness measured at the horizontal plates, becomes close to $T_M$ just outside the thermal BLs and this effect is more pronounced at higher $Ra$. In addition, the BL thickness decreases with increasing $Ra$ and therefore the fraction of the heat current that can avoid the thermal resistance of the fluid by going through the sidewall decreases with increasing $Ra$. {
This statement can be made more quantitative by observing the temperature profiles of figures \ref{fig:figure7}cd (isothermal sidewall) and computing the `temperature defect' as
$D = 1/L\int_0^L | T(z)-T_m |{\rm d}z$ that is a measure of how much the fluid layer next to the sidewall  deviates from the isothermal condition $T(z)=T_m$. We have obtained the values $D=5.0\cdot 10^{-2}$, $3.4\cdot 10^{-2}$, $2.3\cdot 10^{-2}$ and $1.5\cdot 10^{-2}$, respectively, for $Ra=2\cdot 10^7$, $2\cdot 10^8$, $2\cdot 10^9$ and $2\cdot 10^{10}$. The same quantity computed for the profiles of figure \ref{fig:figure7}ab (adiabatic sidewall) yields $D=8.0\cdot 10^{-2}$, $5.2\cdot 10^{-2}$, $3.8\cdot 10^{-2}$ and $2.6\cdot 10^{-2}$ (again for $Ra=2\cdot 10^7$, $2\cdot 10^8$
, $2\cdot 10^9$ and $2\cdot 10^{10}$) confirming that, in this second case, the flow is less isothermal. Nevertheless, being the sidewall perfectly adiabatic, 
no parasite heat currents can be produced through the sidewall.
}

On account of the above scenario nearly the same heat transport is measured in a RB sample with an adiabatic sidewall and an isothermal surface at $T_M$ already when $Ra=2\times10^{10}$.

In the G\"ottingen experiments \cite[]{ahl09} the heat transport measurements have shown a relevant dependence on $T_U-T_m$, where $T_U$ indicates the temperature outside the RB sample and $T_m$ the average fluid temperature in the sample
\footnote{It should be noted that in a laboratory experiment $T_m$ is not { determined} as a volume average of the temperature field but rather as a time average of a pointwise temperature measurement, or a series of measurements, at a vertical position halfway between the plates.}.  It is worth mentioning that it is not at all trivial to decide how an external temperature $T_U$ is "felt" by the `dry' surface of the sidewall because of the complex interaction between the porous convection in the insulating foam and the isothermal surfaces of the various shields (see figure \ref{fig:figure2}). Nevertheless, in a first attempt to model this effect we have changed the temperature of the sidewall to a value different from $T_M$. Figure \ref{fig:figure5}b shows that the heat transport at the bottom and top plate is different when the temperature of the sidewall is different from $T_M$, because then a net heat flux is generated through the sidewall. Using geometrical arguments one can show that the relation between the heat flux at the bottom $Nu_h$ and top $Nu_c$ plate is given by
\begin{equation} \label{Equation_sideflux}
Nu_h+\frac{4}{\Gamma} Nu_{sw}=Nu_c,
\end{equation}
where $Nu_{sw}$ is the heat flux through the sidewall.

\begin{figure}
\centering
\subfigure{\includegraphics[width=0.98\textwidth]{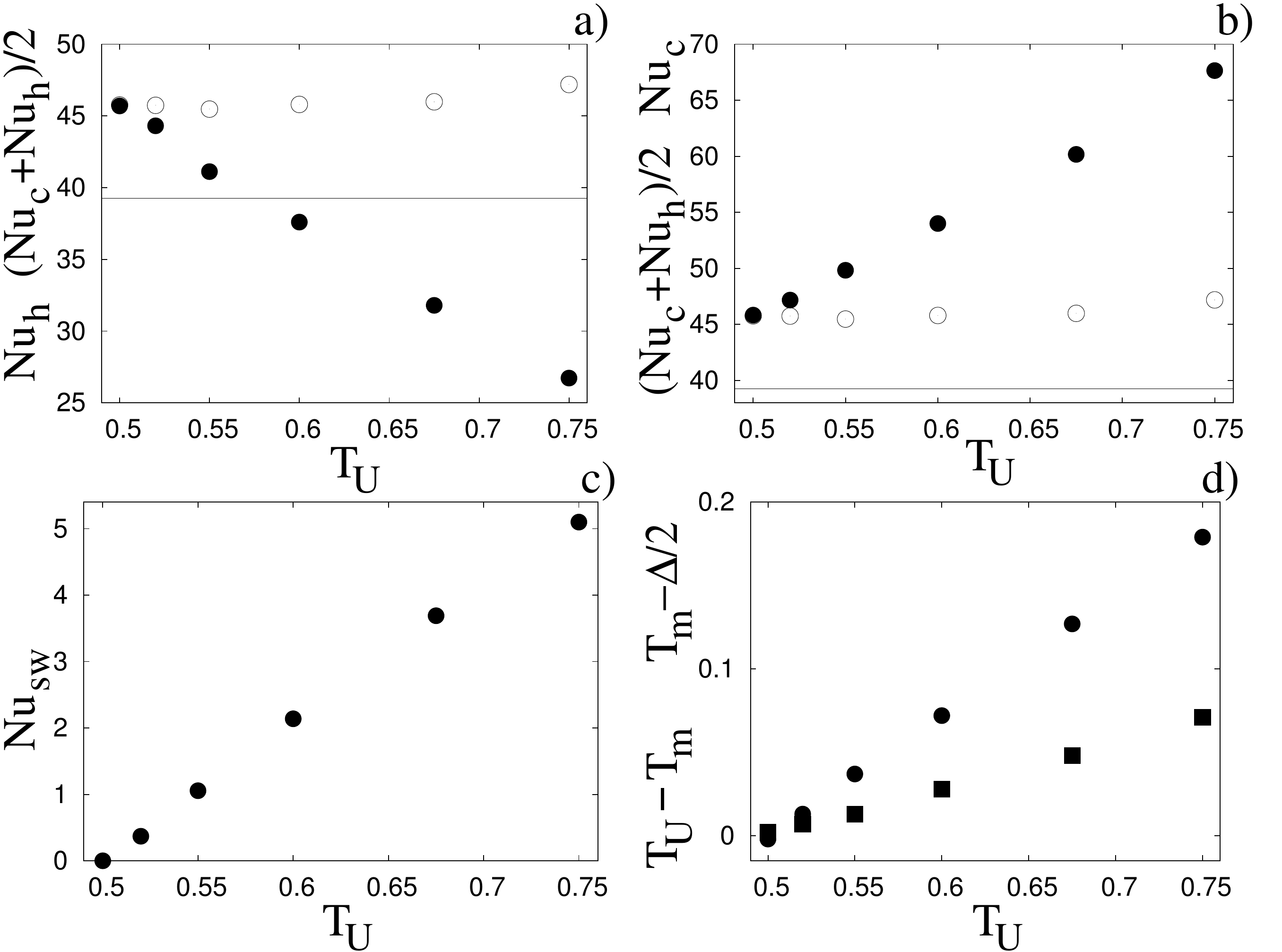}}
\caption{$Nu$ {\it versus} $T_U$ at $Ra=2\times 10^8$: a) at the bottom plate, b) at the top plate (solid circles). Average between top and bottom values (open circles) in a) and b). c) $Nu$ {\it versus} $T_U$ through the sidewall. Panel d) shows the mean temperature $T_m -\Delta/2$ (circles) and $T_U-T_m$ (squares) as function of $T_U$. }
\label{fig:figure8}
\end{figure}

Figure \ref{fig:figure8} shows the time averaged heat flux at the bottom and top plates and through the sidewall as function of the sidewall temperature for $Ra=2\times10^8$ and $Pr=0.7$. The figure shows that increasing the sidewall temperature results in a decrease of the heat transport measured at the bottom plate, but in an increase of the heat flux through the sidewall and the top plate, as is predicted by the relation (\ref{Equation_sideflux}). However, we note that $(Nu_h+Nu_c)/2$ stays approximately constant with $T_U$. This happens because the warmer sidewall warms up the fluid in the lower part of the cell and thereby decreases the heat flux that has to be supplied by the bottom plate. In order to compare this results with the G\"ottingen experiments we need to know $T_U-T_m$ for the different cases. In a first brutal attempt to compare the simulations with the experiments we take for the temperature outside the cell $T_U$ the temperature of the sidewall. Figure \ref{fig:figure8}d shows that $T_U-T_m$ increases when the sidewall temperature is increased. It is worth mentioning that as $Nu_h \neq Nu_c$ it must be decided whether the heat transfer measured at the bottom or at the top plate has to be taken for the comparison. Because in the G\"ottingen experiments \cite{he11} only measured the heat transfer at the bottom plate (via the supplied electrical power) we decided to compare the $Nu$ number measurements with the heat transport at the bottom plate. In agreement with the experiments of \cite{he11} figure \ref{fig:figure6}b shows that a positive $T_U-T_m$ results in a lower heat transport over a wide $Ra$ number range. In this case the heat transport does not converge to the value measured in a cell with an adiabatic sidewall for higher $Ra$, because now the warmer sidewall generates a heat flux through the sidewall, as shown in figure \ref{fig:figure8}c. 

\section{Mixed sidewall boundary conditions} \label{section_Mixed}
As already anticipated before, a completely isothermal sidewall is a too crude oversimplification of the actual experimental configuration. When we examine the design of the G\"ottingen RB sample in figure \ref{fig:figure2}a in more detail we find that there are micro-shields placed just above (below) the bottom (top) plate that are kept at a temperature of $T_M$ in order to prevent the temperature of the plates to influence the isolated region between the sidewall and the side-shield. With the aim of getting closer to the experimental situation we consider the sidewall area adjacent to these micro-shields to have a constant temperature and the rest of the sidewall as adiabatic, since there is a thick insulation layer between the sidewall and the side-shield. Schematically, this configuration is shown in figure \ref{fig:figure3}. The isothermal regions close to the bottom and top plates are kept at a temperature of $T_M$ and have a height of $\Lambda = 0.015L$, which is based on the design of the G\"ottingen RB setup (\cite{he11}).

Figure \ref{fig:figure6}b shows that remarkably the results from this model are almost the same as for the case in which the entire sidewall temperature is kept at $T_M$. The reason is that the fluid temperature close to the sidewall only differs significantly from $T_M$ inside the thermal BLs and most of the heat flux through the sidewall is found in these regions, see figures \ref{fig:figure7}c and d. In fact, when we compare the thickness of the thermal BL with the height of the isothermal region of the sidewall ($\Lambda = 0.015L$) used in the model, we find that already for relatively low $Ra$ the small isothermal sidewall regions are larger than the thermal BL thickness $\lambda_\theta = L/(2Nu)$, which is $0.0188L$ at $Ra=2\times10^7$, $0.0106L$ at $Ra=2\times10^8$, $0.0054L$ at $Ra=2\times10^9$, and $0.0029L$ at $Ra=2\times10^{10}$. Thus in the model the sidewall is isothermal within the thermal BL regions. Therefore the model result is very close to that obtained with a completely isothermal sidewall. We note that in experiments this region of the sidewall that is close to the horizontal plates is particularly challenging to control. In fact, in this region the sidewall and the horizontal plates meet and therefore it is impossible to completely prevent all spurious heat currents and deviations from the intended ideal problem.

\section{Sidewall with physical properties} \label{section_Physical}

In this section we investigate the influence of a sidewall with finite thickness and physical thermal properties on the measured heat transport. There are several different ways to join the sidewall with the hot and cold plates at the bottom ($z=0$) and top ($z=L$) surfaces. Here, in order to simplify the computation and the imposition of the boundary conditions, we have chosen to `extend' the plates also below and above the sidewall (see the sketch of figure \ref{fig:figure3}).
 In this case the sidewall and the bottom and top plate are in direct contact.
It is worth mentioning that this is only one among several different possibilities to couple the sidewall with the plates and, although it might resemble the arrangement of the Oregon/Trieste experiment  (\cite{nie00}) it has been motivated mainly by its computational simplicity. Another possibility could be that the sidewall extends below (above) the hot (cold) plate and surrounds it which is close to the Grenoble and Brno setups (\cite{roc01b}, \cite{urb12}). The latter configuration has not been simulated in order to maintain the total amount of runs to a reasonable number.
Other differences might come from flanges that are placed externally to the cell to join the sidewall with the plates or to 
connect different segments of the former (\cite{nie00}); this would result in an effective wall thickness that is different
from its nominal value. 

\begin{figure}
\centering
\subfigure{\includegraphics[width=0.65\textwidth]{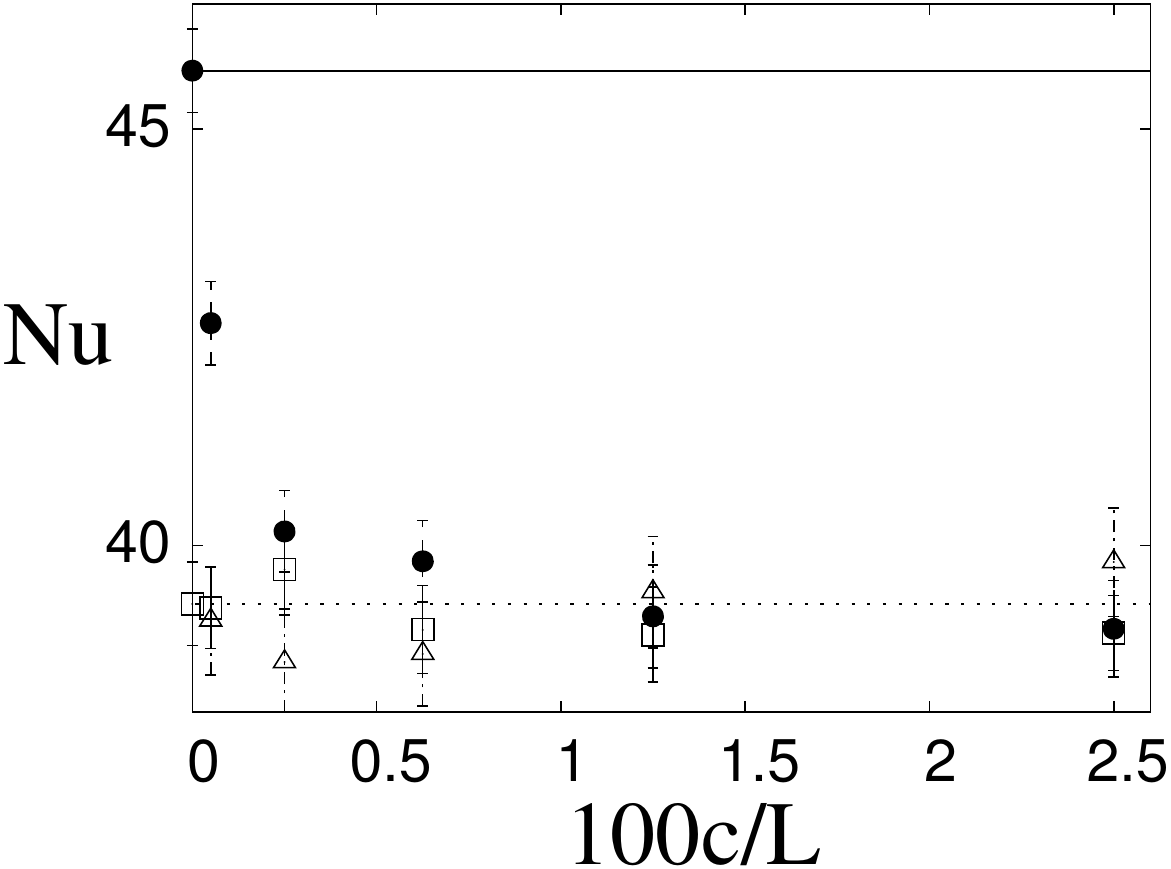}}
\caption{
{
The heat transfer at $Ra=2\times10^8$ in a $\Gamma=1/2$ sample with a stainless steel sidewall and filled 
with a gas at $Pr=0.7$ (open triangles) for adiabatic temperature boundary condition on the `dry' side of the sidewall. 
The other symbols are the data for Plexiglas sidewall and water as working fluid ($Pr=7.0$): solid 
circles and open squares indicate, respectively, the results for an isothermal ($T_M$) and adiabatic temperature 
boundary condition on the `dry' side of the sidewall. The dotted and solid  lines indicate the adiabatic and isothermal ($T_M$) $Nu$ for zero--thickness sidewalls, respectively.}
}
\label{fig:figure9}
\end{figure}

\begin{figure}
\centering
\subfigure{\includegraphics[width=0.92\textwidth]{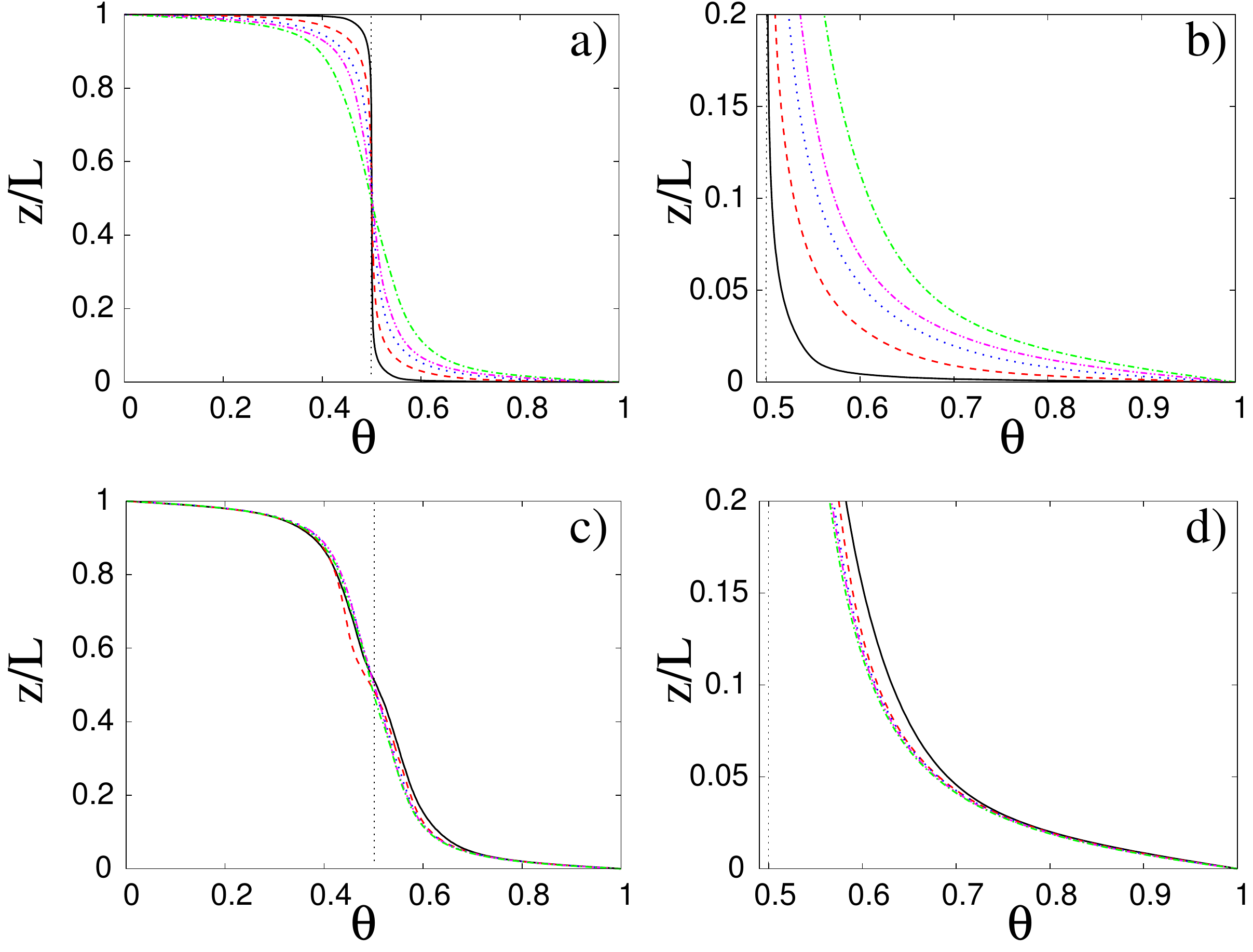}}
\caption{
{
The measured temperature profile at the sidewall/fluid interface as function of the height when a container 
with a Plexiglas sidewall is filled with water ($Pr=7.0$ and $Ra=2\times 10^8$) and the `dry' surface of the sidewall 
is isothermal at temperature $T_M$, a) and b). The colors indicate the different sidewall thicknesses: black $c=0.005L$, red $c=0.0025L$, blue $c=0.0075L$, magenta $c=0.0125L$ and green $c=0.025L$. The plots on the left hand side show the profiles over the entire domain and the plots on the right hand side show only the region close to the horizontal plate. 
c) and d), the same as a) and b) but for adiabatic `dry' surface of the sidewall.}
}
\label{fig:figure10}
\end{figure}

{
We keep the fluid volume constant and vary the thickness of the sidewall by setting $R_w > R_f$, see figure \ref{fig:figure3}. 
Differently from the results of the previous section, where only the Rayleigh and Prandtl numbers account for the 
thermal properties of the system, here we need to specify also the material properties of the sidewall in order to
solve the conjugate heat transfer problem.

Initially, we consider a $\Gamma=1/2$ sample with a stainless steel sidewall filled with gaseous helium (Pr=0.7) at $T_M 
= 4.2$~K. 
The corresponding material properties are $\rho_w/ \rho_f=485$, $C_w/C_{pf}=0.00022$, and $\lambda_w/\lambda_f=44.5$. In this case we compare with results from previous sections only for adiabatic temperature boundary conditions on the `dry' side ($r=R_w$) of the stainless steel sidewall because in cryogenic helium experiments the convection cell is placed in a vacuum.
Figure \ref{fig:figure9} shows that for the cryogenic helium/stainless steel 
combination the heat transport depends weakly the sidewall thickness $c=R_w-R_f$, at least for $Ra=2\times10^8$ and $Pr=0.7$, which is in agreement with the results from \cite{ver02} where this configuration has been analyzed in detail.}
{
We consider further a $\Gamma=1/2$ cell with a Plexiglas sidewall filled with water ($Pr=7.0$), which has been 
adopted in several recent experiments  (\cite{xi08} 
with the material properties $\rho_w/ \rho_f=1.16$, $C_w/C_{pf}=0.239$, and $\lambda_w/\lambda_f=0.344$ ). 
}
For this configuration we use both adiabatic and constant temperature boundary conditions on the `dry' side ($r=R_w$) of the Plexiglas sidewall.
Figure \ref{fig:figure9} shows that the heat transport as function of the sidewall thickness $c$ depends on 
the temperature boundary condition on the `dry' side of the sidewall. 
In particular, when the latter is isothermal, the heat transport decreases for 
increasing wall thickness; of course as $c \rightarrow 0$, $Nu$ recovers the value 
of figure \ref{fig:figure6} for an isothermal zero--thickness sidewall. For 
increasing $c$, owing to the isothermal boundary condition that gets further from 
the fluid/wall interface, $Nu$ decreases and eventually drops slightly below the ideal 
(adiabatic zero--thickness sidewall) value since some of the heat escapes the 
fluid and flows through the sidewall. Figure \ref{fig:figure10} shows that this behavior is related to the temperature of the sidewall at the fluid 
interface. The figure shows that when the sidewall is very thin and the `dry' side 
of the sidewall is maintained at $T_M$ the temperature at the fluid--wall 
interface is very close to $T_M$, except for a very small region close to the 
horizontal plates. This confirms that a very thin sidewall with these physical 
properties is indeed close to the case of a perfect isothermal sidewall. 
However, with increasing sidewall thickness the region in which the temperature of the sidewall at the fluid interface deviates from $T_M$ increases significantly.
 This implies that for increasing $c$ the radial heat--flux through the sidewall is significantly lower than with a perfect isothermal surface and therefore it gets closer to a sample with an adiabatic sidewall.

{
Also in this case it can be noted that when the `dry' surface of the sidewall is adiabatic 
the effect of the sidewall thickness on the heat transfer is much less pronounced than with an 
isothermal boundary condition and it shows negligible sensitivity to the sidewall thickness. The reason is that the temperature profiles at the fluid--wall interface 
and at $r=R_f-\delta_\theta$ are always close enough to prevent significant spurious heat fluxes and, owing to the reduced thermal conductivity of the wall, 
the temperature profiles do not change with $c$. Accordingly
it is observed that the differences in $Nu$ are of the order of $2$--$3\%$  
and comparable to the actual precision of the heat transfer measurements 
($\sim 2\%$) of laboratory experiments. 
}
Nevertheless, as shown in section \ref{section_bimod}, relevant changes in the flow structure can be produced 
by changing the sidewall properties even when the $Nu$ number is relatively unchanged.

%As a further analysis we have simulated a case in which the ambient temperature $T_U$ is different from $T_M$ and its effect on the flow is mediated by the presence of a sidewall with physical properties. We have assumed $T_U = T_c+0.75 \Delta$, $c/L =0.0125$, $Ra=2\times 10^8$, $Pr = 0.7$ and gaseous helium/stainless steel properties. Similarly to the results of figure \ref{fig:figure8} we found strongly different values of the $Nu$ numbers at the hot and cold plates, a significant heat flux through the sidewall, and a bulk temperature different from $T_M$. In particular we have obtained $Nu_c = 57.5$, $Nu_h = 23.4$ and $T_m = 0.680$ 
%(to be compared with $Nu_c = 67.6$, $Nu_h = 26.7$ and $T_m = 0.679$ for the analogous case of figure \ref{fig:figure8}). It is interesting to note that the heat flux from the external ambient to the sidewall through the `dry' side is very large, i.e.\ $Nu_{dry} = 14.3$, but most of it flows through the sidewall without entering the fluid sample. Nevertheless from the above data it is clear that the sidewall alone cannot completely prevent spurious radial heat fluxes and this is the reason why additional thermal shields are commonly used in laboratory experiments.

We wish to point out that the results of figure \ref{fig:figure9} describe a general behavior that holds regardless of the particular fluid/sidewall combinations even though the presented numbers here are specific for a cryogenic gaseous helium/stainless steel  or ambient temperature Plexiglas/water setups. In fact, for the G\"ottingen experiments of \cite{he11} with compressed $SF_6$ and a Plexiglas sidewall ($\rho_w/\rho_f = 11.705$, $C_w/C_{pf} = 2.004$, $\lambda_w/\lambda_f = 13.66$, and using $c/L =0.0125$) { at $Ra=2\times 10^8$ and $Pr=0.7$ we find $Nu=38.6$ using an isothermal boundary condition of $T_M$ at the `dry' side and $Nu=38.6$ using an adiabatic boundary condition at the `dry' side. Note that this is below the value $Nu = 39.5$ computed for the ideal setup.}

\section{Isolation layer} \label{Isolation layer}
All the results described above have been obtained assuming that the ambient temperature boundary conditions can be applied directly at the `dry' side of the sidewall. Although this is already an improvement with respect to the direct imposition of the boundary condition at the fluid/wall interface, the real situation is far more complex because in between the sidewall and the ambient there are usually additional insulating layers and sometimes thermal shields (see figure \ref{fig:figure2}). On the other hand we have already mentioned that the sidewall alone cannot prevent spurious radial heat fluxes from outside when the ambient temperature $T_U$ is different from $T_M$ and for this reason we have simulated also some cases in which the sidewall is covered by an insulating layer of foam and some thermal shields.

\begin{figure}
\centering
\centering
\subfigure{\includegraphics[width=0.65\textwidth]{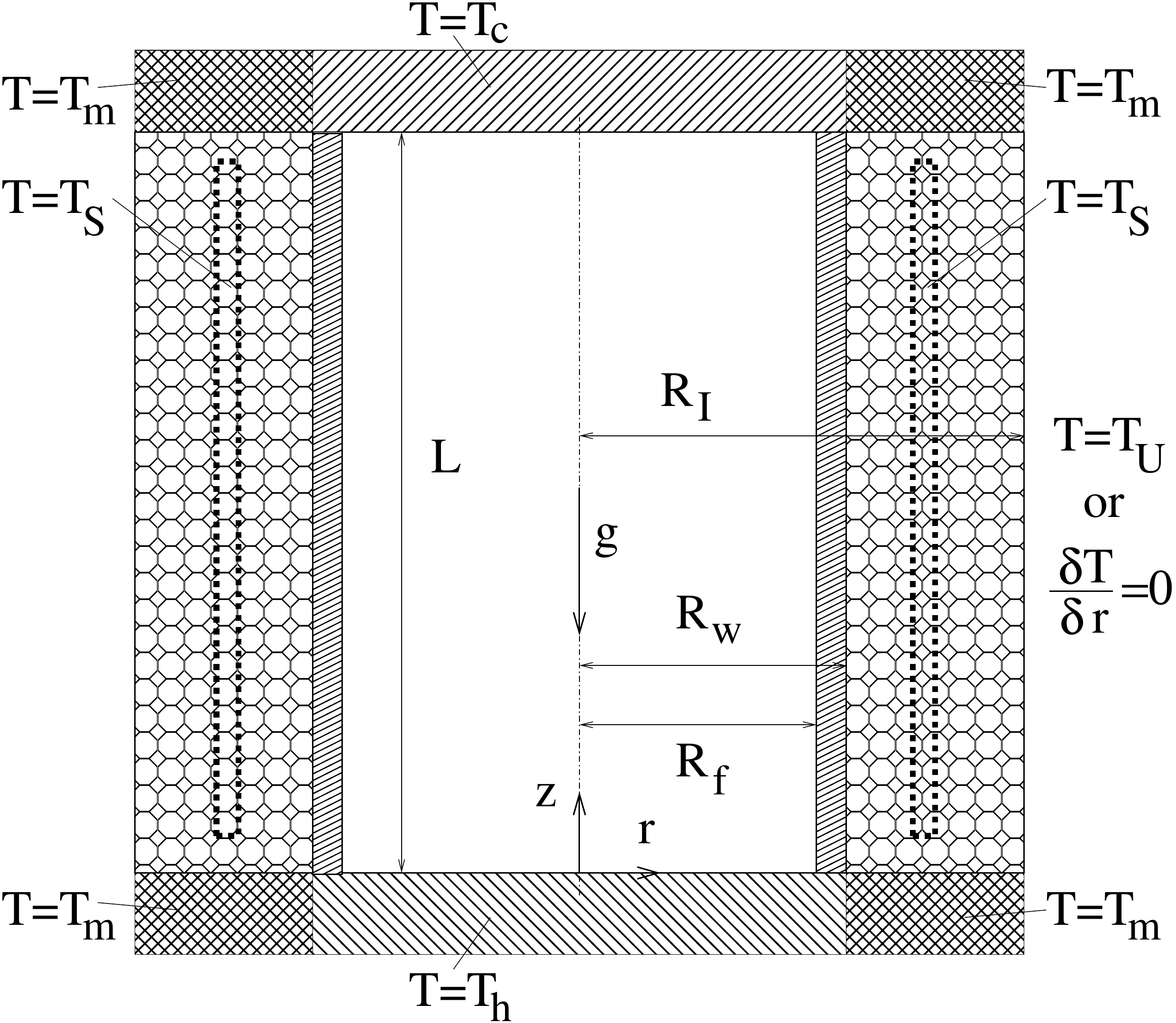}}
\caption{Sketch of the numerical setup for the cell with an external insulating layer and thermal shields: The upper and lower boundaries are isothermal. The sidewall has a thickness $c = R_w-R_f$ and the porous foam is $f=R_I-R_w$ thick; its 'external side' at $r=R_I$ can be either adiabatic $\partial T/\partial r=0$ or isothermal at a temperature $T=T_U$. Within the foam volumes thermal shields with a prescribed temperature can be placed. }
\label{fig:figure11}
\end{figure}

The simulated configuration is sketched in figure \ref{fig:figure11}. The insulating foam (G. Ahlers, {\it Personal Communication}) has been assumed of open--cell type, to prevent its collapse when operating in pressurized environments as in the experiments of \cite{he11}, and therefore porous convection can occur. In order to model also this phenomenon we have resorted to an immersed boundary method \cite[]{fad00} that modifies equation (\ref{eqt1}) to
\begin{eqnarray}
 \frac{D\textbf{u}}{Dt} = - \nabla P + \left( \frac{Pr}{Ra} \right)^{1/2} \nabla^2 \textbf{u} + \theta \textbf{$\widehat{z}$} + {\bf f}
\label{eqt2}
\end{eqnarray}
and allows to handle fluid, solid and porous media with a single equation. More in detail, the forcing term ${\bf f}$ assumes a different expression depending on the particular point in the domain:
\begin{enumerate}[i)]
\item ${\bf f =0}$ in the fluid so that equation (\ref{eqt2}) reduces to the Navier--Stokes equation,
\item ${\bf f}$ has a value that ensures that ${\bf u=0}$ within the solid parts (sidewall and thermal shields, see \cite{fad00} for more details),
\item ${\bf f} = -{\bf u}/K$ in the insulating foam to allow for porous convection with pressure losses that depend on the porosity $K$ (Navier--Stokes--Brinkman equation).
\end{enumerate}
Note that equation (\ref{eqt}) remains unmodified even if $\rho$, $C$ and $\lambda$ assume the value of the foam or the shields when the point is in those media.

Looking at figure \ref{fig:figure11}, and considering different combinations of materials, thicknesses of walls and layers in addition to multiple shields at various positions, it is immediately clear that a complete {parameter} study is almost impossible owing to the enormous number of possible configurations. We have therefore considered only two cases, one with a layer of foam and without shields and another with three shields arranged as in \cite{he11}.
For these simulations we have have used the same resolutions as the previous cases at $Ra=2\times 10^8$ in the vertical and azimuthal direction. In contrast, the mesh in the radial direction had a larger number of radial nodes to simulate also the phenomena in the foam layer and in the thermal shields and the computational points were non-uniformly distributed (by a third order spline) to capture 
the boundary layers at the interfaces. Finally, the mesh had
$193\times 131\times 257$ nodes in the azimuthal, radial and vertical directions (while the previous cases at $Ra=2\times 10^8$ were run on meshes of $193\times 85\times 257$ nodes).

\begin{figure}
\centering
\centering
\subfigure{\includegraphics[width=0.92\textwidth]{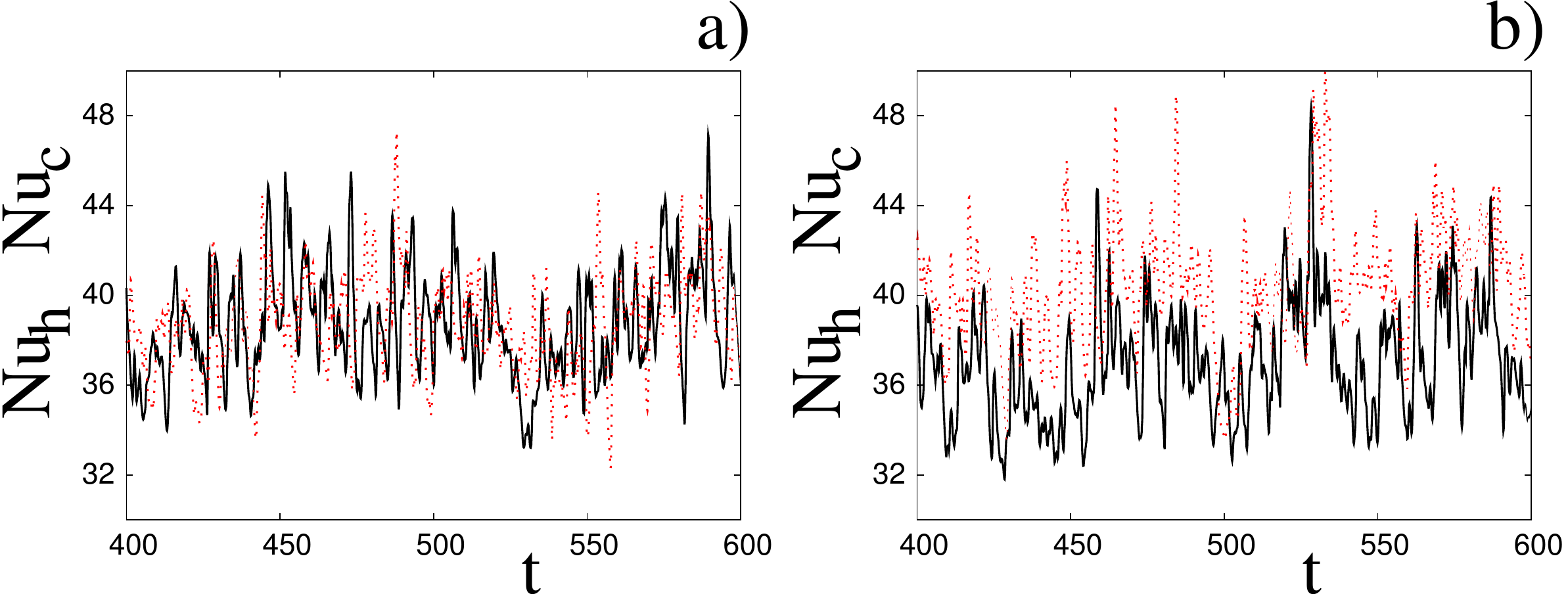}}
\caption{(Color online) Time evolution of the $Nu$ numbers at the hot (\solid and black) and cold (\dashed and red) plates for a flow at $Ra=2\times 10^8$ using compressed $SF_6$ ($Pr=0.7$) with Plexiglas sidewall of thickness $c=0.0125L$ and a layer of insulating foam of thickness $R_I-R_w = 0.1375 L$: a) ambient temperature $T_U = T_M$, b) ambient temperature $T_U = T_c + 0.55\Delta$.}
\label{fig:figure12}
\end{figure}

For the first case we have assumed a foam layer of thickness $R_I-R_w = 0.1375 L$ and with properties $\rho_I= 2\rho_f$, $C_I = C_{pf}$ and $\lambda_I = 5 \lambda_f$ { and a Plexiglas sidewall with thickness $c=0.0125L$}. For the non--dimensional porosity we have used the value $K = 10$ after having verified by preliminary simulations that the order of magnitude of the velocities within the foam was about fifty times smaller than that in the fluid. { Figure \ref{fig:figure12}a shows the resulting $Nu$ using an ambient temperature $T_U = T_M$ are consistent with the value of $Nu=38.6$ found while modeling just the Plexiglas sidewall.  Figure \ref{fig:figure12}b shows that when $T_U = T_c + 0.55\Delta$ gives $Nu_c = 40.1$ and $Nu_h = 37.1$ and both values are smaller than those of figure \ref{fig:figure8} for a similar $T_U$ but without a sidewall with physical properties.} Nevertheless the fact that $Nu_c > Nu_h$ suggests that, despite the layer of foam, some heat flux is entering the fluid through the sidewall and indeed it is confirmed by a direct computation. It is worth mentioning that this behavior shows up only after a very long initial transient, of the order of $\sim 10^3$ time units at $Ra=2\times 10^8$, since it takes a long time before the system with a thick insulating foam layer { reaches} the thermal equilibrium. Running similar cases at higher $Ra$ is therefore unfeasible because the length of the initial transient will increase with increasing $Ra$, while the time step becomes smaller and the simulation cost per time step higher.

\begin{figure}
\centering
\centering
\subfigure{\includegraphics[width=1.05\textwidth]{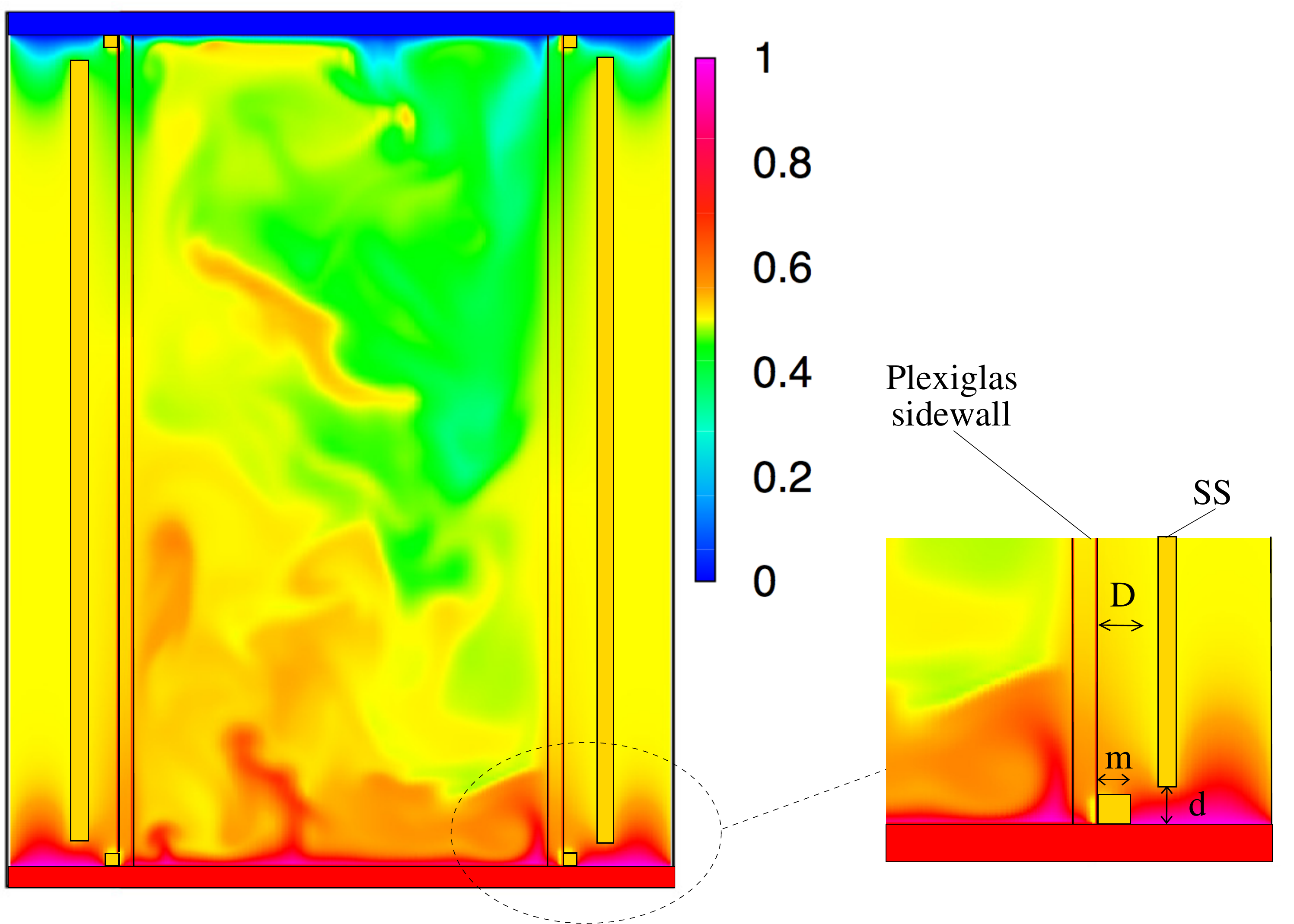}}
\caption{Instantaneous snapshot of the temperature field at $Ra=2\times 10^8$ and $Pr=0.7$ for a setup with compressed $SF_6$ and a Plexiglas sidewall of thickness $c=0.01L$. The ambient temperature is $T_U = T_c + 0.5\Delta$. This setup reproduces the micro thermal shields BMS and TMS, toroidal rings of square cross section $m\times m$ with $m=0.015L$, at temperature $T_M$ and the side-shield SS, also at $T_M$. The side-shield has a thickness of $0.01L$, a distance from the `dry' side of the sidewall equal to $D=0.03L$ and a distance from the bottom and top boundaries of $d=0.015L$. This setup is inspired from the G\"ottingen experiment sketched in figure \ref{fig:figure2}. Note that in this particular setup, the portion of the upper and lower surfaces ($z=0$ and $z=L$) outside the sidewall ($r\geq R_w$) are set at the temperature $T_M$. On the right there is a detail of the sidewall, plate and shields junction. Note that the lower plate is conventionally 
{ colored} red while the fluid at the highest temperature $\theta = 1$ is { indicated} by magenta.   }
\label{fig:figure13}
\end{figure}

Figure \ref{fig:figure13} shows an instantaneous temperature snapshot for a flow at $Ra=2\times 10^8$ and $Pr=0.7$ in a setup inspired by, but not identical to, the G\"ottingen experiment of \cite{he11} with compressed $SF_6$ and a Plexiglas sidewall of thickness $c=0.01L$. The ambient temperature is $T_U = T_c + 0.5\Delta = T_M$ and the setup includes also the top and bottom thermal micro-shields (BMS) and (TMS), torii of square cross section $0.015L\times 0.015L$ at temperature $T_M$, and a side-shield (SS) of thickness $0.01L$ at temperature $T_M$ (see figure \ref{fig:figure2}). For this configuration we have obtained a bulk temperature $T_m = 0.500$ that is indistinguishable from $T_M$ and also the $Nu$ numbers $Nu_c = 39.4$ and $Nu_h = 39.5$ agree with $Nu = 39.5$, which is measured in the `ideal' RB cell with adiabatic sidewalls. 

It is interesting to note that nearly identical results for the Nusselt numbers and bulk temperature have been obtained in a setup as in figure \ref{fig:figure13} but with the surfaces at $z=0$ and $z=L$ beyond the sidewall ($r> R_w$) at a constant temperature $T_M$. 

The arrangement of thermal shields and isolation layers of figure \ref{fig:figure13} turned out to be very effective to prevent
the effects of the external ambient temperature on the flow. In an additional simulation, in fact, the external temperature 
was set to $T_U = T_c + 0.675\Delta$ obtaining a bulk temperature $T_m = 0.499$ and the Nusselt numbers $Nu_c = 38.5$ and 
$Nu_h = 38.4$ that, though both smaller, agree within the statistical error ($\sim 3-4\%$) with the reference value of $Nu = 39.5$.

It should be noted however that this result is extremely sensitive to the sealing between the cell and the thermal shields.
In fact, in the setup of figure \ref{fig:figure13} the boundaries at $z=0$ and $z=L$ are no--slip also beyond the sidewall ($r> R_w$)
and the thermal shield $SS$ extends vertically up to a distance of $0.015L$ from the horizontal plates. 
As a result the foam in the volume in between the sidewall and the thermal side-shield is almost closed. Hence the velocities in the foam are more than hundred times smaller than in the fluid. 
In another simulation we have reduced the vertical length of the side-shield so that it extended in between $0.06L \leq z \leq 0.94L$ instead of the range $0.015L \leq z \leq 0.985L$ of the previous case. Due to this small change in the shield extension the bulk temperature slightly raised to $T_m = 0.507$, { with an uncertainty below $1\%$}, which is in the same direction as the results of figure \ref{fig:figure8}d. The $Nu$ numbers, however, remained equal to the ideal $Nu$ within the statistical uncertainty ($\sim 3-4\%$) and further conclusions can not be drawn. 

Despite the effort in setting up a numerical simulation close to the  G\"ottingen laboratory experiment of \cite{he11}  there are still 
 details that make the two samples slightly different: In the latter case the volume in between the sidewall and the shield $SS$ is
 sealed by a film of lexan. In addition, the direct contact between the top and bottom plates and the micro-shields (BMS, TMS)
is prevented by another layer of lexan (G. Ahlers {\it Personal Communication}). Both details have 
not been included in the numerical simulations.

Based on the above results it is clear that the flow can be influenced by the details of the experimental apparatus {\it outside} the fluid region. Since the possible combinations of shields, thicknesses, positions, materials and temperature boundary conditions is too large to be covered even by an {\it ad hoc} investigation, it would be advisable to run simulations of specific cases in order to test in advance a particular geometry. Nevertheless, as a general conclusion it can be said that the junction between the horizontal plates and the sidewall is particularly critical and also the use of thermal shields and their positioning, though generally beneficial, should be carefully considered.

\section{Sidewall effects on the flow} \label{section_bimod}
In all the previous sections we focussed on how the sidewall affects the heat transfer. In this section we show that the sidewall properties can also influence the flow structures and that this does not necessary have to be reflected in the $Nu$ number.

\begin{figure}
\centering
\subfigure{\includegraphics[width=0.99\textwidth]{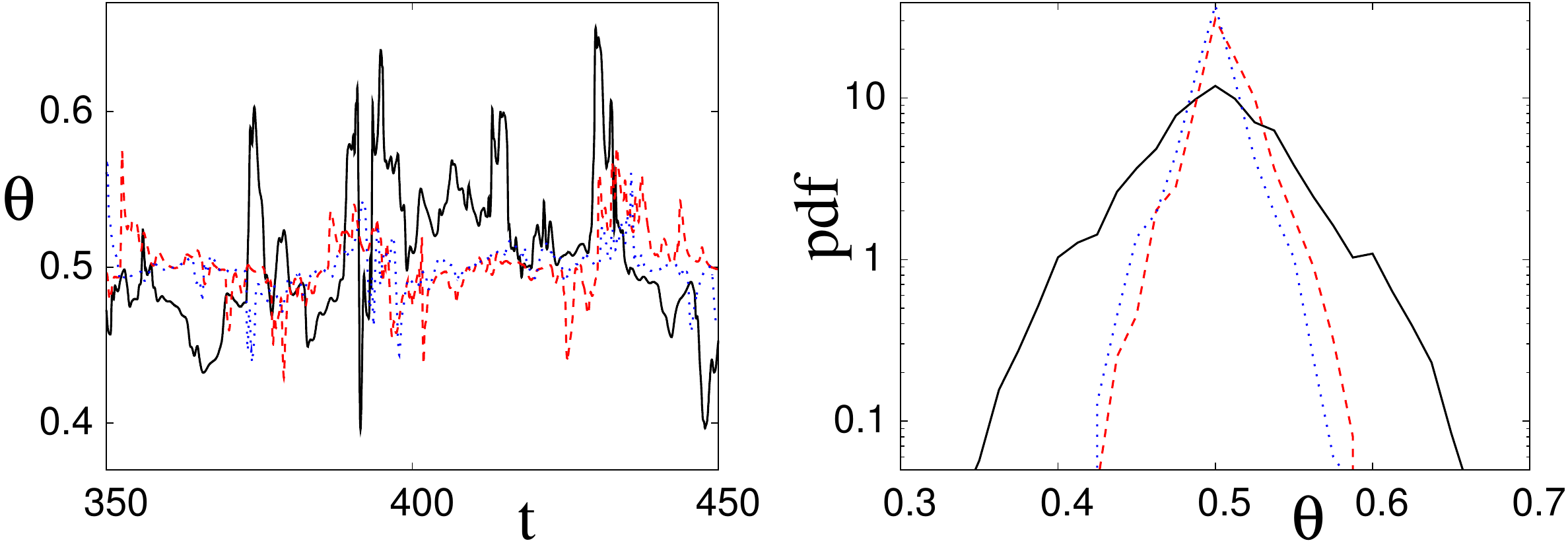}}
\caption{
{
(Color online) a) Temperature time series sampled at $z=0.5L$, $r=R_f-\delta_\theta$ (with $\delta_\theta = L/(2Nu)$), and $\phi=0$ at $Ra=2\times 10^8$ and $Pr=0.7$: \solid (black) adiabatic sidewall, \dashed (red) isothermal sidewall and \dotted (blue) sidewall of finite thickness with isothermal `dry' side ($c=0.0125L$), Plexiglas/pressurized $SF_6$ combination). b) Histograms of the time series of panel a). In the adiabatic case the fluctuations are much larger.} 
}
\label{fig:figure14}
\end{figure}

From figure \ref{fig:figure10} it is already evident that a different sidewall temperature boundary condition can change the mean temperature profiles in the nearby flow region. However, these changes are only significant close to the horizontal plates where most of the (spurious) radial heat flux occurs. The situation is quite different for the temperature fluctuations since an adiabatic boundary allows for any fluctuation while an isothermal surface tends to anchor the fluid temperature to its own value. This behavior is evidenced in figure \ref{fig:figure14} where temperature time series sampled by a probe at mid-height ($z=L/2$) and at the same radial distance from the boundary ($r=R_f-\delta_\theta$) are shown for an adiabatic, isothermal, and finite--thickness sidewall. The first evident difference for the first two cases is that, although both temperatures are fluctuating about the mean value $T_M$, the fluctuations are smaller for the constant temperature sidewall, which is consistent with the above conjecture. 
{
The case with finite--thickness sidewall, with its own physical properties, shows fluctuations that are of the same order as those of the isothermal boundary even if the latter is not directly in contact with the fluid. In this case, however, the heat dynamics in the solid wall is coupled with that in the fluid and many different behaviours can be obtained by changing the solid and fluid properties. 
}

We wish to stress, however, that this latter result depends on the temperature boundary condition on the `dry' surface of the sidewall, on its thickness and on its physical properties through the ratios $\lambda_w/\lambda_f$, $\rho_w C_w/(\rho_f C_{pf})$ (see equation \ref{eqt}). As a matter of fact by varying the sidewall properties and its temperature boundary conditions it is possible to obtain, for the flow region next to the sidewall, any behavior ranging from the absence of temperature fluctuations up to the maximum for the ideal adiabatic boundary with zero heat capacity.

\begin{figure}
\centering
\subfigure{\includegraphics[width=0.49\textwidth]{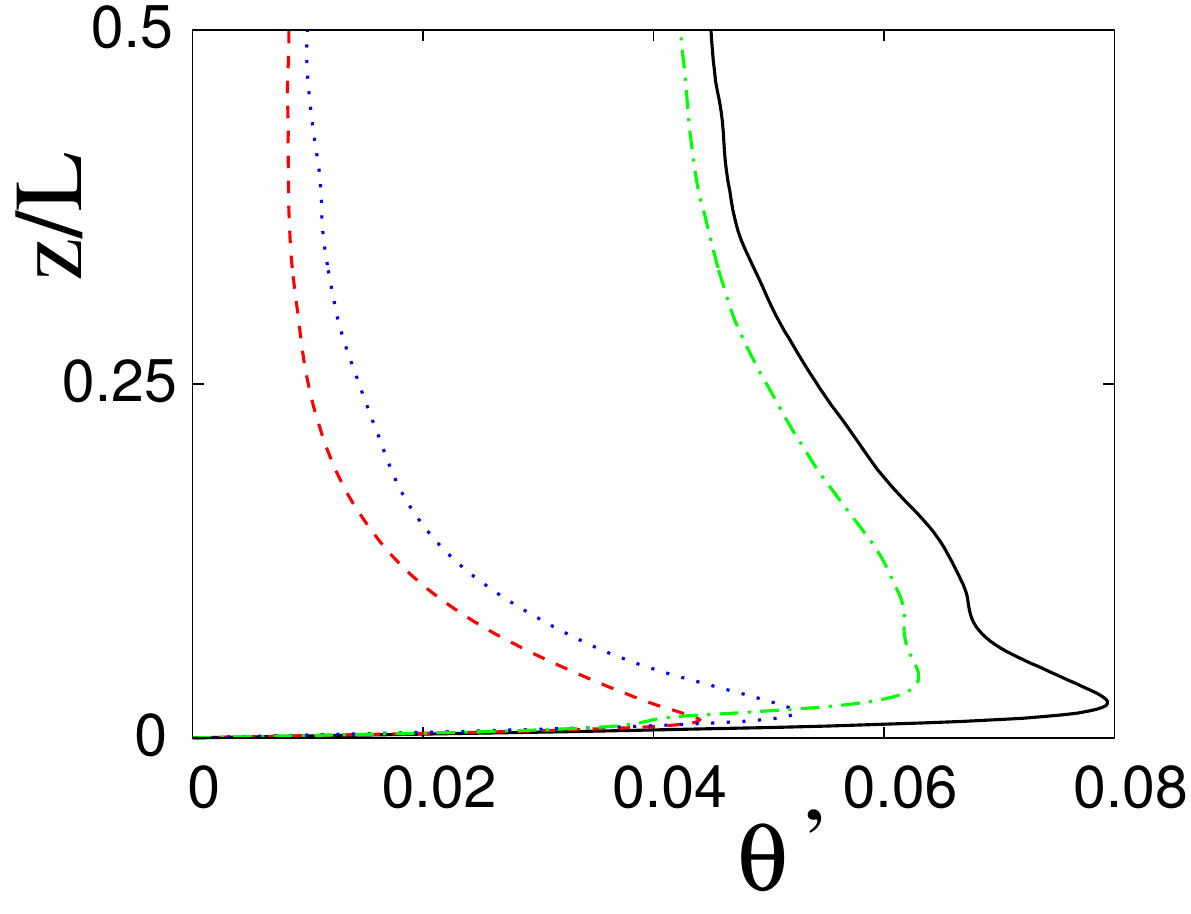}}
\caption{
{
(Color online) Time and azimuthally averaged vertical rms temperature profiles at a distance $\delta_\theta = L/(2Nu)$ from the `wet' surface of the sidewall at $Ra=2\times 10^8$ and $Pr=0.7$: \solid (black) ideal adiabatic sidewall, \dashed (red) isothermal sidewall, \chndot (green) mixed boundary conditions as in section \ref{section_Mixed} and \dotted (blue) sidewall of finite thickness ($c= 0.0125L$ Plexiglas/pressurized $SF_6$ combination) with isothermal temperature boundary condition on the `dry' side.} 
}
\label{fig:figure15}
\end{figure}

The { analysis in this section} was motivated by the study of \cite{ahl12} showing that the experimental measurements of wall--close vertical temperature profiles in the range $8\times 10^{12} \leq Ra \leq 10^{15}$ behaved according to a logarithmic law even in the low end of $Ra$ where it was not expected since they did not belong to the ultimate regime predicted by \cite{kra62} and \cite{gro11}.
 The data of the numerical simulations from \cite{ste10d} at $Ra=2\times 10^{12}$ not only confirmed the logarithmic temperature profiles but showed excellent agreement with the experimental fits \cite{ahl12}. In contrast, the {\it temperature fluctuation} profiles in experiments and simulations, though both could be fitted by a logarithmic law did not agree: In the numerical simulations the fluctuations were about five times larger than in the laboratory measurements. In a successive analysis part of the disagreement was found to be caused by the too large thermal inertia of the thermistors that acted as a low--pass filter on the temperature fluctuations (G. Ahlers {\it Personal Communication}). Another part of the mismatch, however, should be ascribed to the different nature of the sidewall that, in the numerical simulation is ideal and adiabatic, therefore allowing for the maximal fluctuations, while in the experiment it is of finite thickness and made of Plexiglas. Figure \ref{fig:figure15} shows that indeed the sidewall can alter the profiles of the temperature fluctuations already at $Ra=2\times 10^8$ and that a wide range of possible behaviors can be obtained by changing the sidewall properties and the temperature boundary condition on the `dry' side.

\begin{figure}
\centering
\centering
\subfigure{\includegraphics[width=1.05\textwidth]{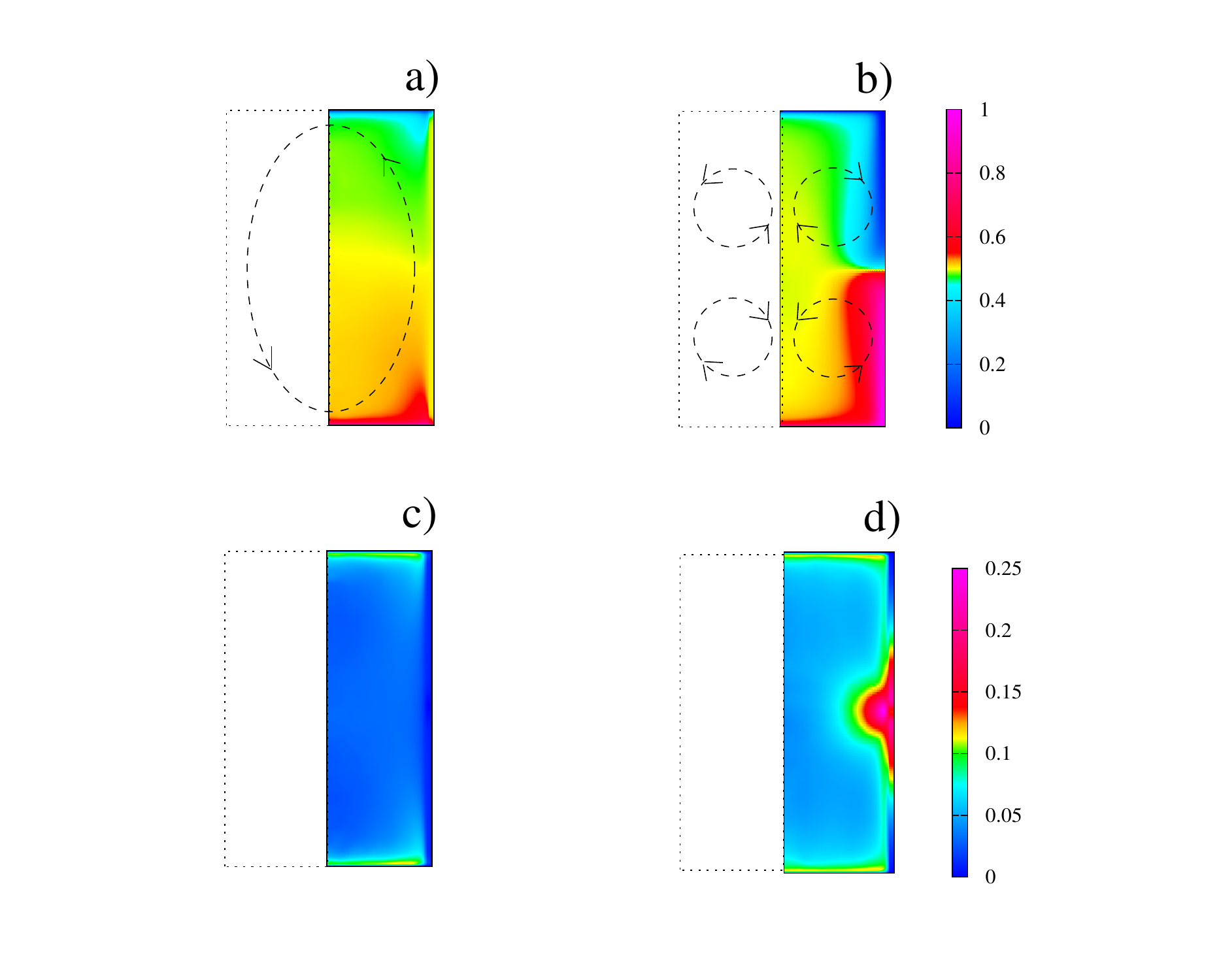}}
\caption{
{
Time and azimuthally averaged temperature, (a) and (b), and rms temperature fluctuations, (c) and (d), for a flow at $Ra=2\times 10^8$ and $Pr=0.7$ for a setup with finite thickness sidewall ($c=0.0125L$) with
$\rho_wC_w/(\rho_f C_{pf})=2925$, and $\lambda_w/\lambda_f=1000$. The panels a) and c) are for an isothermal `dry' surface of the sidewall, the panels b) and d) for an adiabatic temperature boundary condition. Note that the time-- and azimuthal--average produces a single r--z meridional plane, the other half of the section is sketched by a dotted line for clarity. In panels a) and b) also a drawing of the mean flow structure is shown.} 
}
\label{fig:figure16}
\end{figure}

{
Before concluding this section we show that the sidewall can also play a key role in determining the mean flow structure. 
To this aim have simulated an {\it artificial} example in which a gas is bounded on the side by a very conductive sidewall
($\rho_wC_w/(\rho_f C_{pf})=2925$, and $\lambda_w/\lambda_f=1000$) of thickness $c=0.0125L$.
The thermal conductivity ratio is clearly exaggerated and this value is unlikely in a real experimental apparatus; here, however,
we want to stress the effects on the mean flow structure of adiabatic and isothermal boundary conditions on the `dry' 
surface of the sidewall and these are enhanced by a highly conductive material.

In figure \ref{fig:figure16} we show the mean and rms temperature maps at $Ra=2\times 10^8$ and $Pr=0.7$ for two cases, with adiabatic and isothermal `dry' sidewall. In the first case the mean flow consists of two vertically stacked counter rotating toroidal vortices, while for the isothermal `dry' boundary condition the large scale is the classical single roll state. Figure \ref{fig:figure16}c and d show that these flows produce completely different temperature fluctuations as argued at the beginning of this section.
}

Clearly, the reason for the generation of two torii of figure \ref{fig:figure16}b is the undesired temperature distribution along the sidewall that extends the plates also along the vertical boundary, therefore forcing the flow also from the side. In contrast, when the `dry' sidewall is forced to be isothermal at temperature $T_M$, the heat flowing from the horizontal plates to the sidewall can escape from the system directly through the isothermal surface without entering the fluid. 

We wish to stress that the structures of figure \ref{fig:figure16}a and b are not hysteretic configurations resulting from a particular initial condition but rather stable states to which the system relaxes. As a check, we have used the standard single--roll configuration of figure \ref{fig:figure16}a as initial condition for the setup of figure \ref{fig:figure16}b and we have verified that after a very long transient (of about $800L/U$ time units) the flow undergoes a slow adjustment and eventually it recovers the two--torii configuration of figure \ref{fig:figure16}b. It is worth mentioning that despite the very different mean flow in both cases of figure \ref{fig:figure16} the bulk temperature was $T_m = T_M$ and the differences of the $Nu$ numbers were well within those of isothermal and adiabatic sidewall temperature boundary conditions.

{ Finally, in this section, just as in other cases analyzed in this paper}, the thickness of the sidewall has been exaggerated in order to make its effect already evident at $Ra=2 \times 10^8$ when the mean flow is intense and it dominates the flow dynamics. We find that, as $Ra$ increases the large--scale circulation weakens and therefore it is likely that also thinner sidewalls with reduced heat capacity would be able to force a particular flow state.

\section{Summary} \label{section_Conclusion}
We used direct numerical simulations (DNSs) to investigate the influence of the physical properties and the temperature boundary conditions of the sidewall on the heat transport in Rayleigh-B\'enard (RB) convection. The cases we considered are inspired by the experiments of \cite{ahl09d}, { \cite{xi08}, and \cite{nie00}}. 
 \cite{he11} found two different branches for the heat transport: A slightly higher heat transport is measured when $T_U-T_m$ is negative, where $T_U$ is the temperature outside the cell and $T_m$ is the average fluid temperature, and vice versa when $T_U-T_m$ is positive.

We show that keeping the temperature of the sidewall fixed at $T_M$ leads to a higher heat transport at lower $Ra$, because part of the heat current circumvents the thermal resistance of the fluid by going through the sidewall. However, this effect disappears at higher $Ra$ where the bulk becomes more isothermal and the heat flux through the sidewall decreases. In agreement with the experimental results we find that an increase of the sidewall temperature, with respect to the value $T_M$, leads to a lower heat transfer at the bottom plate. Just as in experiments this effect is visible over a large $Ra$ number regime; at least up to $ Ra = 2 \times 10^{10}$, there are no indications that the effect decreases for increasing $Ra$. Subsequently, we argue that in the G\"ottingen RB setup the temperature boundary condition should be close to adiabatic in the center region and close to a constant temperature condition of $T_M$ in the vicinity of the horizontal plates due to the use of the micro-shields. The heat transfer we measure at the bottom plate in this model is the same as the heat transfer that is obtained when the entire sidewall is kept at $T_M$. This shows that the sidewall region close to the horizontal plates is crucial, while this is particularly challenging region in experiments as the sidewall and the horizontal plates meet in this region and therefore it is impossible to complete prevent all heat currents in that region. 

The flow dynamics is more complex when the sidewall is considered with its thermal properties and thickness 
{
because the heat dynamics in the fluid couples with that in the wall.
For RB samples filled with water ($Pr=7.0$) and with a Plexiglas lateral boundary, the Nusselt number shows little sensitivity to the sidewall thickness 
when its `dry' surface is adiabatic.
In contrast, when the `dry' side of the sidewall is isothermal,
 the temperature at the fluid--wall interface becomes such that large heat fluxes through the sidewall are generated. However these spurious fluxes decrese
as the wall thickness increases and, for thick enough sidewalls 
($c \geq 0.0075L$) the deviations of $Nu$ are below $2$--$3\%$.
}

As a proof--of-concept, resorting to the Navier--Stokes--Brinkman equations, we have simulated a case with an additional external insulation layer of foam, where porous convection can occur, and some others with the foam and several thermal shields. The former showed that, apart from producing very long transients, the foam alone could not avoid the flow to be affected by an ambient temperature $T_U$ different from $T_M$. In contrast, the combined use of foam and thermal shields adequately prevented undesired effects. Nevertheless, it has also been shown that the shielding of the region next to the junction between plates and sidewall is really crucial and small changes in the shield position are sensed by the flow already at $Ra=2\times 10^8$.

We wish to point out that all the simulations with the foam and the shields were performed at $Ra=2\times 10^8$ 
owing to the augmented complexity of the problem that largely increased the computational time of the simulations. Although
the results have given useful indications about some additional effects they have been obtained for a Rayleigh number that
is few orders of magnitude below the range of the experiments ($Ra=10^{11}$--$10^{15}$). The present results, therefore, should not
be trivially applied to the experiments without further considerations on very high Rayleigh number flows.

Finally, it was shown that the sidewall not only affects the heat transfer but can also influence the temperature fluctuations and even the mean flow structure. Owing to the very large number of parameters   that can influence the flow it is unfeasible to explore the complete phase space. By analyzing some particular aspects, however, we hope to have shed some light on details that should be kept under control when designing a new setup or running an experiment.\\

\noindent {\it Acknowledgement:} We thank Guenter Ahlers for stimulating discussion and for providing data and details of his experiments. The presented simulations were performed on Huygens (DEISA (Distributed European Infrastructure for Supercomputing Applications) project), CASPUR (Inter-University Consortium for the Application of Super-Computing for Universities and Research), and HLRS (High Performance Computing Center Stuttgart). We gratefully acknowledge the support of Wim Rijks (SARA) and we thank the DEISA Consortium (www.deisa.eu), co-funded through the EU FP7 project RI-222919, for support within the DEISA Extreme Computing Initiative. RJAMS was financially supported by the Foundation for Fundamental Research on Matter (FOM).

%\bibliographystyle{/Users/stevensrjam/Documents/work/PhD/Biblatex/jfm}
%\bibliography{/Users/stevensrjam/Documents/work/PhD/Biblatex/literatur}
%\newpage

\appendix
\section{Tables with the results of the simulations}

In this Appendix we report the key data of the simulations performed in the paper; Table \ref{Table Sim1} contains the data for the
zero--thickness sidewall (see section \ref{section_isothermal} and section \ref{section_Mixed}). Table \ref{Table Sim2} summarizes the simulations with the finite thickness sidewall with thermal properties of section \ref{section_Physical}.
 
\begin{table}
  \centering
\begin{tabular}{|c|l|c|c|c|c|}
  \hline
  $Ra$ & SW ($T_U$) & $T_m$ & $Nu_{h}$ & $Nu_{c}$ & $Nu_{sw}$  \\
  \hline
  $2 \times 10^6$ & adiab. & $0.50$   & 10.63          & 10.64        & $\approx$ 0  \\
  $2 \times 10^6$ & isot. ($0.5$) & $0.50$   & 16.73          & 16.74        & $\approx$ 0 \\  \hline
  $2 \times 10^7$ & adiab. & $0.50$   & 20.50          &  20.42       & $\approx$ 0  \\
  $2 \times 10^7$ & isot. ($0.5$) & $0.50$   & 26.41          &26.40        & $\approx$ 0  \\
  $2 \times 10^7$ & isot. ($0.75$) & $0.68$   & 15.86          & 39.10        & 3.01\\ 
  $2 \times 10^7$ & mixed  & $0.50$   & 28.03          & 28.00        & $\approx$ 0  \\ \hline
  $2 \times 10^8$ & adiab. & $0.50$   & 39.50          &  39.60       & $\approx$ 0  \\
  $2 \times 10^8$ & isot. ($0.5$) & $0.50$   & 45.70          &45.83       & $\approx$ 0  \\
  $2 \times 10^8$ & isot. ($0.52$) & $0.51$   & 44.31          &47.13       & 0.37  \\
  $2 \times 10^8$ & isot. ($0.55$) & $0.54$   & 41.12          &49.83       & 1.02  \\
  $2 \times 10^8$ & isot. ($0.60$) & $0.57$   & 37.60          &54.00       & 2.14  \\
  $2 \times 10^8$ & isot. ($0.675$) & $0.63$   & 31.80          &60.17       & 3.69  \\
  $2 \times 10^8$ & isot. ($0.75$) & $0.68$   & 26.73          & 67.65        & 5.04  \\
  $2 \times 10^8$ & mixed  & $0.50$   & 47.22          & 46.96        & $\approx$ 0  \\ \hline
  $2 \times 10^9$ & adiab. & $0.50$   & 79.75          &  79.73       & $\approx$ 0  \\
  $2 \times 10^9$ & isot. ($0.5$) & $0.50$   & 89.36          &89.31       & $\approx$ 0  \\
  $2 \times 10^9$ & isot. ($0.75$) & $0.67$   & 54.99          &131.51      & 10.20  \\
  $2 \times 10^9$ & mixed  & $0.50$   & 90.56          & 91.21        & $\approx$ 0  \\ \hline
  $2 \times 10^{10}$ & adiab. & $0.50$   & 173.10          & 173.48       & $\approx$ 0  \\
  $2 \times 10^{10}$ & isot. ($0.5$) & $0.50$   & 171.58          & 171.16       & $\approx$ 0  \\
  $2 \times 10^{10}$ & isot. ($0.75$) & $0.68$   & 102.28          & 277.51       & 22.41  \\
  $2 \times 10^{10}$ & mixed & $0.50$   & 173.84          &  173.33      & $\approx$ 0  \\
  \hline
\end{tabular}
  \caption{Summary of the simulations performed for the configuration with `zero thickness' sidewall: 
The columns from left to right indicate the Rayleigh number ($Ra$), the sidewall temperature boundary condition, adiabatic (adiab.) or isothermal (isot.) and for the latter the imposed temperature ($T_U$). 
$T_m$ is the volume averaged fluid temperature, $Nu_h$ and $Nu_c$, respectively, the Nusselt numbers computed as surface averages at the hot and cold plates. $Nu_{sw}$ is the Nusselt number evaluated as surface average on the sidewall.
All the simulations are performed at $Pr=0.7$. }
\label{Table Sim1}
\end{table}
  
Following the papers by \cite{ahl00,roc01b,ver02,nie03} we can compute the total heat flowing through the hot plate as
$Q_T = \int_0^{2\pi} \int_0^{R_w} \lambda \nabla \theta \cdot {\bf n} {\rm d}S$ and the heat going from the hot plate directly 
into the sidewall as $Q_w = \int_0^{2\pi} \int_{R_f}^{R_w} \lambda \nabla \theta \cdot {\bf n} {\rm d}S$, where $\lambda$ is either
the thermal conductivity of the fluid $\lambda_f$ or of the sidewall $\lambda_w$ depending on the point of the plate if it is in
contact with the former or the latter. The heat entering the fluid layer is trivially $Q_f = Q_T-Q_w$ and it should be
used to compute the  Nusselt number. In laboratory experiments, only $Q_T$ is available and, according to \cite{ahl00},
 the factors $f_w = Q_w/Q_T$ or $f_f = Q_f/Q_T$, if
known from some model, could be used to correct the measured quantity $Q_T$ to compute a corrected Nusselt number via $Nu_{corr} = 
f_fQ_T/S$. 

In Table \ref{Table Sim2} we report the factor $f_f$ as obtained by the present numerical simulations in which $Q_T$ and $Q_f$
could be computed separately. We note that the case for cryogenic helium and $c/L = 0.0025$ agrees with the results of
\cite{ver02}.
The values obtained for the stainless steel/gaseous helium setup imply larger differences than those for the combination Plexiglas/water: This is not surprising on account of the bigger ratio of the thermal conductivities $\lambda_w/\lambda_f$ of the former case. 
For the combination of Plexiglas and water the correction is bigger for isothermal boundary conditions on the `dry' side of the
sidewall than for the adiabatic case.

We also report the wall number $W =(4/\Gamma)(\lambda_w/\lambda_f)(c/L)$ and the correction factor $F = 1/[1+f(W)]$ with
$Nu_{corr} = FQ_T/S$ being
$f(W) = [A^2/(\Gamma Nu)](\sqrt{1+2W\Gamma Nu/A^2}-1)$ as defined by \cite{roc01b} with $A = 0.8$. 

Finally we emphasize that the Nusselt number even if computed by $Nu = Q_f/S$ deviates substantially from the value $Nu_{ideal}$ as it comes from a
simulation with zero--thickness adiabatic sidewall in the case of isothermal sidewall boundary condition. Therefore, even if $f$ 
would be given by some reliable model it would not return the ideal Nusselt number owing to the changes produced in the flow by the 
conjugate heat transfer between the lateral boundary and the fluid layer. 

\begin{table}
  \centering
\begin{tabular}{|c|c|c|c|l|c|c|c|c|c|c|}
  \hline
  $Pr$ & $ (\rho C)_w/(\rho C_p)_f$ & $\lambda_w/\lambda_f$ & $100c/L$ & SW ($T_U$) & $T_m$ & $Nu_{h}$ & $Nu_{c}$ & $f_f$ 
  & $W$ & $F$ \\
  \hline
   $0.7$ & $0.107$ & $44.5$ & $0.05$ & adiab. & $0.50$   & 39.04   & 39.22    & 0.964   & 0.178 & 0.926\\
   $0.7$ & $0.107$ & $44.5$ & $0.25$ & adiab. & $0.50$   & 38.70   & 38.51    & 0.877   & 0.89  & 0.827\\
   $0.7$ & $0.107$ & $44.5$ & $0.625$ & adiab. & $0.50$   & 38.82   & 38.51   & 0.784   & 2.225 & 0.743 \\
   $0.7$ & $0.107$ & $44.5$ & $1.25$ & adiab. & $0.50$   & 39.56   & 39.32    & 0.594   & 4.450 & 0.665 \\
   $0.7$ & $0.107$ & $44.5$ & $2.5$ & adiab. & $0.50$   & 39.68   & 39.95     & 0.210   & 8.900 & 0.580 \\ \hline
   $7.0$ & $0.277$ & $0.344$ & $0.05$ & adiab. & $0.50$   & 39.37   & 39.24   & 0.999   & 0.0014& 0.999   \\
   $7.0$ & $0.277$ & $0.344$ & $0.25$ & adiab. & $0.50$   & 38.08   & 38.46   & 0.998 & 0.007 & 0.993   \\
   $7.0$ & $0.277$ & $0.344$ & $0.625$ & adiab. & $0.50$   & 39.13   & 38.80  & 0.992   & 0.017 & 0.986    \\
   $7.0$ & $0.277$ & $0.344$ & $1.25$ & adiab. & $0.50$   & 39.06   & 38.91   & 0.909 & 0.034 & 0.976  \\
   $7.0$ & $0.277$ & $0.344$ & $2.5$ & adiab. & $0.50$   & 39.16   & 38.75    & 0.968  & 0.069 & 0.960 \\
   $7.0$ & $0.277$ & $0.344$ & $0.05$ & isot. ($0.5$)& $0.50$   & 42.73   & 42.61 & 0.970 &0.0014 & 0.999     \\
   $7.0$ & $0.277$ & $0.344$ & $0.25$ & isot. ($0.5$)& $0.50$   & 40.25   & 40.12 & 0.938 & 0.007 & 0.997     \\
   $7.0$ & $0.277$ & $0.344$ & $0.625$ & isot. ($0.5$)& $0.50$   & 39.73   & 39.89 & 0.919 & 0.017 & 0.986      \\
   $7.0$ & $0.277$ & $0.344$ & $1.25$ & isot. ($0.5$)& $0.50$   & 39.26   & 39.08  & 0.927 & 0.034 & 0.976    \\
   $7.0$ & $0.277$ & $0.344$ & $2.5$ & isot. ($0.5$)& $0.50$   & 38.86   & 39.15 & 0.891  & 0.069 & 0.960    \\ \hline
   $0.7$ & $23.45$ & $13.66$ & $1.25$ & adiab. & $0.50$   & 38.71   & 38.55      & 0.899  & 1.366 & 0.790\\
   $0.7$ & $23.45$ & $13.66$ & $1.25$ & isot. ($0.5$)& $0.50$   & 38.54   & 38.67 & 0.213  & 1.366 & 0.790   \\
  \hline
\end{tabular}
  \caption{Summary of the simulations performed for the configuration with `finite thickness' sidewall with thermal properties: 
The columns from left to right indicate the Prandtl ($Pr$) numbers, the ratio of the specific heat capacities 
($\rho C$) of wall and fluid, the ratio of their thermal conductivities $\lambda$, the sidewall thickness, the `dry surface' sidewall temperature boundary condition, adiabatic (adiab.) or isothermal (isot.) and for the latter the imposed temperature ($T_U$). 
$T_m$ is the volume averaged fluid temperature, $Nu_h$ and $Nu_c$, respectively, the Nusselt numbers computed as surface 
averages at the hot and cold plates. $f_f$ is the ratio of the heat flowing through the sidewall ($Q_w$) and the total heat ($Q_T$)
imposed to the system. $W =(4/\Gamma)(\lambda_w/\lambda_f)(c/L)$ is the wall number and  $F = 1/[1+f(W)]$ with 
$f(W) = [A^2/(\Gamma Nu)](\sqrt{1+2W\Gamma Nu/A^2}-1)$ as defined by \cite{roc01b}.
All the simulations are performed at $Ra=2 \times 10^8$. 
}\label{Table Sim2}
\end{table}

\end{document}